\documentclass[structuredastract]{aa} 
\usepackage{graphicx, amsmath, xspace,txfonts}
\usepackage{longtable, lscape}
\def\lsim{\mathrel{\rlap{\lower4pt\hbox{\hskip1pt$\sim$}}
    \raise1pt\hbox{$<$}}} 
\def\gsim{\mathrel{\rlap{\lower4pt\hbox{\hskip1pt$\sim$}}
    \raise1pt\hbox{$>$}}} 
\def\kms{$\mbox{km~s}^{-1}$\xspace}
\def\lam{$\lambda$}
\def\lamlam{$\lambda\lambda$}
\def\Ang{$\text{\AA}$\xspace}

\begin{document}

\title{The circumstellar environment and evolutionary state of the 
supergiant B[e] star
Wd1-9\thanks{This work is based on observations collected at the European
  Southern Observatory, Paranal (programme IDs ESO 087.D-0355, 087.D-0440, 087.D-0673,  and 073.D-0327) and uses the ISO-SWS database of Sloan et al. (2003).}}

   \author{J.S.~Clark\inst{1} \and B.~W.~Ritchie\inst{1} \and I.~Negueruela\inst{2}}
   \offprints{J.S.~Clark, \email{s.clark@open.ac.uk}}

   \institute{
        Department of Physics and Astronomy, The Open University, Walton Hall, 
        Milton Keynes MK7 6AA, United Kingdom
   \and 
        Departamento de F\'{\i}sica, Ingenier\'{\i}a de Sistemas y 
        Teor\'{\i}a de la Se\~{n}al, Universidad de Alicante,
        Apdo. 99, 03080 Alicante, Spain
   }

   \date{Accepted ??? Received ???}

   \abstract
   {Historically, supergiant (sg)B[e] stars have been difficult to include in  theoretical schemes for the evolution of massive OB 
stars.}
{The  location of Wd1-9 within the coeval starburst cluster Westerlund 1 means that
 it may be placed into a proper evolutionary context and we therefore aim to utilise a comprehensive 
multiwavelength dataset to determine its physical properties and consequently its 
relation to other sgB[e] stars and the global population of massive evolved stars within Wd1.}  
  {Multi-epoch R- and I-band VLT/UVES and VLT/FORS2 spectra are used to constrain the properties of the circumstellar 
gas, while an ISO-SWS spectrum covering
    2.45--45$\mu$m is used to investigate the distribution, geometry and
    composition of the dust via a semi-analytic irradiated
    disk model. Radio emission enables a long term mass-loss history to be determined, while X-ray 
observations reveal the physical nature of high energy processes within the system.}
{Wd1-9 exhibits the rich optical emission line spectrum that is characteristic of sgB[e] stars. Likewise its 
 mid-IR spectrum resembles those of the LMC sgB[e] stars R66 and 126, revealing
 the presence of equatorially concentrated silicate dust, with a mass of
  $\sim$10$^{-4}M_{\odot}$. Extreme historical and ongoing mass loss 
($\gtrsim 10^{-4} M_{\odot}$yr$^{-1}$)
is inferred from the radio observations. The  X-ray properties of Wd1-9  imply the presence of high temperature plasma
within the system and are directly comparable to a number of confirmed
short-period colliding wind binaries within Wd1.}
  {The most complete explanation for the observational properties of Wd1-9 is that it is a massive interacting 
binary currently undergoing, or recently exited from, rapid Roche-lobe overflow, supporting the hypothesis that 
binarity mediates the formation of (a subset of) sgB[e] stars.
 The mass loss rate of Wd1-9 is  consistent with such an assertion, while viable progenitor 
and descendent systems are present within  Wd1 and comparable sgB[e] binaries have been identified in the Galaxy. 
Moreover, the rarity of sgB[e] stars - only two examples are identified from a  census of $\sim68$  young massive Galactic clusters
 and associations containing $\sim600$ post-Main Sequence stars - is explicable given the rapidity ($\sim 10^4$yr) expected 
for this phase of massive binary evolution.}

  \keywords{stars: evolution - stars: emission line - circumstellar
    matter - stars: individual: Wd1-9} 

  \titlerunning{The sgB[e] binary Wd1-9} \maketitle
%

\section{Introduction}

\begin{figure*}
\begin{center}
\resizebox{\hsize}{!}{\includegraphics{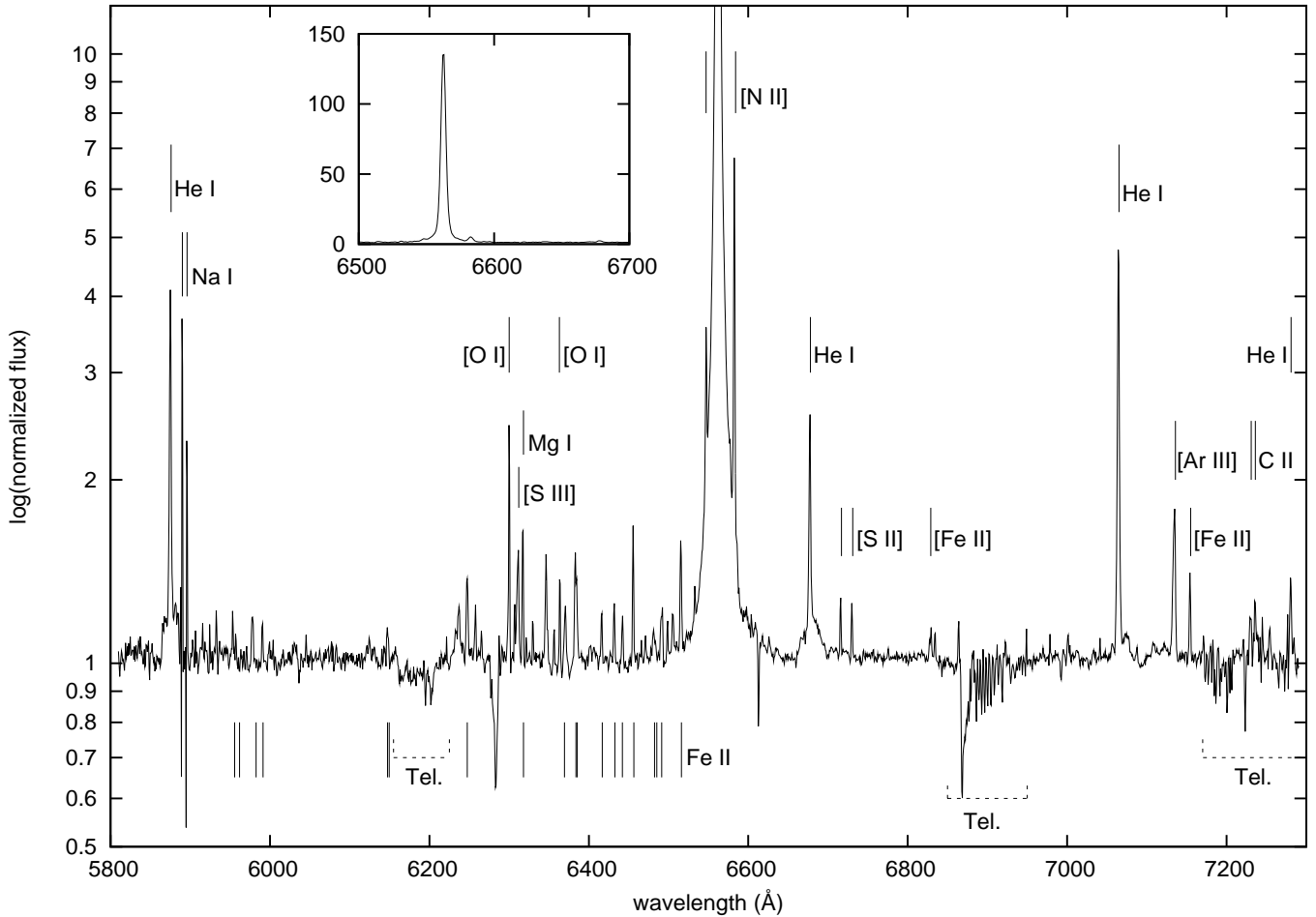}}
\caption{Semi-logarithmic VLT/FORS2 spectrum from 2004 covering
  $5800-7300\text{\Ang}$. Rest wavelengths of the principal emission
  lines are indicated, and telluric bands are also identified. The
  low-resolution inset shows a linear plot of the
  $6500-6700\text{\Ang}$ region to highlight the strength of the
  H$\alpha$ emission.}
\label{fig:rband_spectrum}
\end{center}
\end{figure*}

\begin{figure*}
\begin{center}
\resizebox{\hsize}{!}{\includegraphics{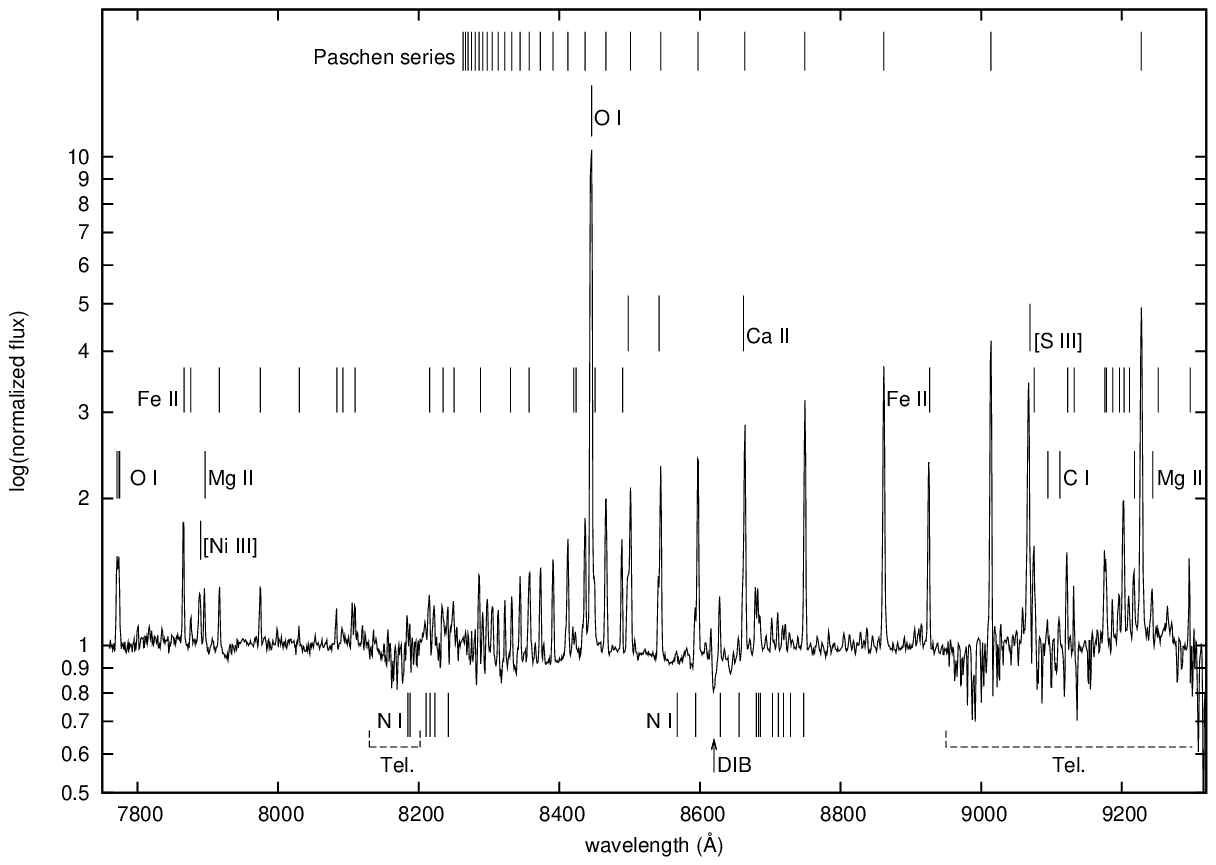}}
\caption{Semi-logarithmic VLT/FORS2 spectrum from 2004 covering
  $7750-9300\text{\Ang}$. Rest wavelengths of the principal emission
  lines are indicated, and telluric bands are also identified. Note
  that O\,{\sc i}~\lam8446 is saturated in this spectrum; the line
  reaches $\sim$35$\times$ the continuum level in unsaturated spectra.}
\label{fig:iband_spectrum}
\end{center}
\end{figure*}

Supergiant (sg)B[e] stars are luminous, evolved objects that
display the \textit{B[e] phenomenon} -- strong Balmer-series emission
lines, permitted emission lines from low-ionization metals and  forbidden
lines of [Fe~II] and [O~I] in the optical spectrum and
 a strong infra-red excess from hot dust (Lamers et  al. \cite{lamers}). These features
are thought to originate in a non-spherical geometry in which the gaseous and dusty circumstellar
material is concentrated in equatorial regions, while
broad UV resonance absorption lines indicate a fast, low-density polar
wind (Zickgraf et al. \cite{zick85}).  Notwithstanding  these uniform classification criteria, 
pronounced differences in both spectral morphologies (e.g. Zickgraf et al.
\cite{zick03}) and    the quantity and temperature of circumstellar dust (cf. Kastner et al. \cite{kastner}, 
\cite{kastner10}, Graus et al. \cite{graus})    are present in sgB[e] stars, which also span  
an unexpectedly wide  range of intrinsic
luminosities (log$(L_{\mathrm{bol}}/L_{\odot}) \sim 4-6$; Gummersbach et al. \cite{gummersbach}).

The physical mechanism(s) driving the  mass loss that sculpts the complex 
circumstellar environments of sgB[e]
stars is unclear at present (Hillier et al. \cite{hillier}) and its 
elucidation is hampered by the intrinsic rarity of 
such stars (Sect. 6.1.2). Luminous blue variables (LBVs) and 
sgB[e] stars are co-located in the 
Hertzsprung-Russell diagram and display  spectroscopic
similarities, leading to suggestions of an evolutionary link
(e.g. Stothers \& Chin \cite{st96}), while the bluewards evolution and current
asymmetric wind of the peculiar cool hypergiant \object{IRC~+10~420}
implies a possible relationship  between B[e] supergiants and 
post-red supergiants (RSG; Davies et al. \cite{davies}).  
More recently, much interest has focused on the role of binarity in triggering the B[e] phenomenon, either 
via merger (e.g. Podsiadlowski et al. \cite{pod}) or   binary driven mass loss (e.g. 
Zickgraf et al. \cite{zick03}, Kastner et al. \cite{kastner});
 we return to this hypothesis in more detail in Sect. 6.

In this paper we examine the multiwavelength properties of the
sgB[e] star Wd1-9 (=\textit{Ara~C} in the notation of Borgman et al. \cite{bks70}), 
in the starburst cluster \object{Westerlund~1}
(hereafter Wd1; Westerlund \cite{w61}).  Classified as a Be star by Westerlund (\cite{w87})
and a sgB[e] star by Clark et al. (\cite{cncg05}), \object{Wd1-9} is radio, mid-IR and X-ray bright 
 (Clark et al. \cite{clark98}, \cite{clark08}, Dougherty et al. \cite{dougherty}) and  displays a rich
emission line spectrum with a complete absence of any photospheric
absorption features. The evolved state of \object{Wd1-9} is firmly
established by its membership of Wd1, which hosts a unique coeval
population of massive, post-main sequence (MS) stars that include
Wolf-Rayets (WRs), numerous OB supergiants, four RSGs and a
population of ten B5--F8 hypergiants with initial masses
$\sim35-40M_{\odot}$ (Ritchie et al. \cite{ritchie10}, Negueruela et al. \cite{negueruela10}). In this regard
 the radio observations of Wd1-9  are of particular interest, 
indicating a recent phase of enhanced or eruptive mass loss, characteristic of either an LBV or interacting binary
nature (Dougherty et al. \cite{dougherty}, Sect. 6).

The structure of the paper is as follows. In Section~\ref{sec:obs_data} we briefly describe our
observations and data reduction, while in Section~\ref{sec:optical} we
describe the optical spectrum in detail. Section~\ref{sec:ir}
describes the IR properties  of \object{Wd1-9}. In
Section 5  we discuss \object{Wd1-9} in the context of other sgB[e] stars and
in Section 6 we  consider its evolutionary state before summarising our findings
in Section 7.

\section{Observations \& data reduction} \label{sec:obs_data}

\begin{table*}
\caption{Journal of observations.}
\label{tab:journal}
\begin{center}
\begin{tabular}{ll|llll}
Date             & MJD$^a$    & Instrument & Mode                     & Wavelength range                  & Reference                   \\
\hline
\hline
24/06/1981         & 44779  & ESO3.6m/B\&C & - & $\sim 6500$ -- 8700\Ang & Westerlund (\cite{w87}) \\   
06/02/1998       & 50850  & ISO/SWS    &  AOT1, speed 1           & 2.38 -- 45.20$\mu$m          & This work          \\
24/06/2001       & 52084  & ESO 1.52m + B\&C & Loral\#38, GRAT\#1 & 6000 -- 10500\Ang  &  Clark  et al. (\cite{cncg05}) \\
07/06/2002       & 52432 & NTT/EMMI   & REMD (2x2) + GRAT\#7     & 6310 -- 7835\Ang  & Clark et al. (\cite{cncg05})\\
06/06/2003       & 52796 & NTT/EMMI   & REMD (2x2) + GRAT\#6     & 8225 -- 8900\Ang  & Negueruela \& Clark 
(\cite{negueruela05}) \\ 
12/06/2004       & 53168  & VLT/FORS2  & longslit (0.3") + G1200R & 5750 -- 7310\Ang & Negueruela et al. (\cite{negueruela10}) \\
13/06/2004       & 53169 & VLT/FORS2  & longslit (0.3") + G1200R & 5750 -- 7310\Ang  &  Negueruela et al. (\cite{negueruela10}) \\
14/06/2004       & 53170  & VLT/FORS2  & longslit (0.3") + G1028z & 7730 -- 9480\Ang  &  Negueruela et al. (\cite{negueruela10})\\
14/06/2004       & 53170  & VLT/FORS2  & longslit (0.3") + G1200R & 5750 -- 7310\Ang  &  Negueruela et al. (\cite{negueruela10})\\
11/03/2006       & 53805  & VLT/ISAAC  & SWS(MR)                  & 22490 -- 23730\Ang  & Mengel \& Tacconi-Garman  \\
                 &        &            &                          &                                 & (\cite{mengel}) \\
13/06/2009         & 54969  & Magellan Clay/MIKE & 0.7" slit & 3200 -- 9000\Ang  & Cottaar et al. (\cite{cottaar}) \\
10/07/2009         & 55023  & Magellan Clay/MIKE & 0.7" slit & 3200 -- 9000\Ang  & Cottaar et al. (\cite{cottaar}) \\
10/07/2010         & 55388  & Magellan Clay/MIKE & 0.7" slit & 3200 -- 9000\Ang  & Cottaar et al.  (\cite{cottaar}) \\

06/04/2011       & 55657 & VLT/UVES   & CD\#4 (860) + IMSL1      & 6708 -- 10426\Ang  & This work\\
14/04/2011       & 55665  & VLT/UVES   & CD\#4 (860) + IMSL1      & 6708 -- 10426\Ang  & This work \\
16/04/2011       & 55667  & VLT/FORS2  & longslit (0.3") + G1028z & 7730 -- 9480\Ang  & This work\\
16/04/2011       & 55667  & VLT/FORS2  & longslit (0.3") + G1200R & 5750 -- 7310\Ang  & This work \\
17/04/2011       & 55668  & VLT/FLAMES & GIRAFFE + HR21           & 8484 -- 9001\Ang  & This work \\
25/04/2011       & 55676  & VLT/UVES   & CD\#4 (760) + IMSL1      & 5708 -- 9464 \Ang  & This work \\
22/05/2011       & 55703 & VLT/FLAMES & GIRAFFE + HR21           & 8484 -- 9001\Ang & This work \\
24/06/2011       & 55736  & VLT/FLAMES & GIRAFFE + HR21           & 8484 -- 9001\Ang & This work \\
\hline
\end{tabular}
\end{center}
$^a$Modified Julian date at the start of the integration.\\
\end{table*}

\object{Wd1-9} was subject to an intensive VLT observing programme during
 2011 April, with dates and instrument configurations listed in
Table~1. High resolution R-~and~I-band spectra were
obtained in service mode on three nights using the Ultraviolet and
Visual Echelle Spectrograph (UVES; Dekker et al. \cite{detal00}) located on VLT
UT2 \emph{Kueyen} at Cerro Paranal, Chile. On the first two nights
(April 6 and 14) UVES was used with the red arm,
cross-disperser CD\#4 and a central wavelength of 8600\Ang, while on the
third night (April 25) dichoric~2 mode was used, with the red
arm using cross-disperser CD\#4 and a central wavelength of 7600\Ang. On
all three nights image slicer \#1 was used, giving a resolving power
$R$$\sim$60,000, with the combination of red and dichoric modes
providing continual spectral coverage from below 6000\Ang to
10420\Ang. Spectra were reduced using the UVES
pipeline\footnote{http://www.eso.org/sci/software/pipelines}, version
4.9.5 and the Common Pipeline Library, version 5.3.1, while final
extraction, correction for heliocentric motion and data analysis were
carried out using the IRAF\footnote{IRAF is distributed by the
  National Optical Astronomy Observatories, which are operated by the
  Association of Universities for Research in Astronomy, Inc., under
  cooperative agreement with the National Science Foundation.}
\textit{onedspec} tasks.

In order to search for longer-term evolutionary trends in the spectrum
of \object{Wd1-9}, we made use of additional intermediate resolution
spectra obtained with the Fibre Large Array Multi Element Spectrograph
(FLAMES; Pasquini et al. \cite{pasq02}), again located on VLT UT2 \textit{Kueyen},
and the spectro-imager FORS2 (Appenzeller et al. \cite{app}), located on VLT UT1
\emph{Antu}. FLAMES spectra were obtained on 2011  April 17,  May 22 and
 June 24 using the HR21 setup, which provides
coverage from 8484--9001\Ang with $R$$\sim$16200; data reduction is as
described in Ritchie et al. (\cite{ritchie09a}).  Six FORS2 spectra 
acquired on 2011 April 16 and 2004 
June 12, 13 and 14 were also utilised.  On all occasions FORS2 was used in
longslit mode ($0{\farcs}3$ slit) with grisms G1200R (5750--7310\Ang) and
G1028z (7730--9480\Ang), giving $R$$\sim$7,000; the FORS2
configuration and data reduction is described in
Negueruela et al. (\cite{negueruela10}). While these data are of lower 
resolution than the
main UVES dataset, the FLAMES spectra allow examination of strong
hydrogen, calcium and nitrogen lines in the months after the UVES
observations, while the FORS2 spectra use identical configurations and
therefore provide a seven-year baseline unaffected by differences in
instrumentation or resolution ( as well as offering an additional opportunity
to search for short term variability in 2004).

We supplemented these observations with 3 epochs of spectroscopy  
obtained with the Magellan Inamori Kyocera
echelle spectrograph in 2009 June and July and 2010 July  (see Cottaar et al.
\cite{cottaar}) that  have 
  similar resolving power to the  UVES data. Additional, lower resolution
data  include two NTT/EMMI spectra from 2002 and 2003  presented and described in Clark 
et al. (\cite{cncg05}) and Negueruela \& Clark (\cite{negueruela05}) respectively, a single spectrum from the ESO 1.52m
 (Clark et al. \cite{cncg05})
and the original  low signal to noise classification spectrum of Westerlund (\cite{w87}) which was obtained in 
1981 and hence extends the baseline of observations to three decades.
Finally, we   make use of the K-band spectrum presented by Mengel \& Tacconi-Garman (\cite{mengel}; Table 1) obtained with 
the VLT/ISAAC on 2006 March 11 and, at longer wavelengths, a mid-IR spectrum
 obtained with the Infrared Space Observatory (ISO)
short wavelength spectrometer (SWS; de Graauw et al. \cite{degraauw}) on 1998 February 6 and taken from
the database of Sloan et al. (\cite{sloan}).

\section{Spectroscopy}\label{sec:optical}

\object{Wd1-9} displays a very rich R-~and~I-band ($\sim 5800$ --  9300\Ang)
emission line spectrum (Figs. 1 and 2) with a complete absence of any apparent photospheric
absorption lines; the only absorption lines visible in the spectrum of
\object{Wd1-9} are interstellar lines of Na~I and K~I, strong diffuse
interstellar bands at 6281, 6614, 8620 and 8648\Ang
(cf. Negueruela et al. \cite{negueruela10}), and telluric bands. Along with very
strong emission lines of H, He\,{\sc i} and O\,{\sc i} \lam8446, there are many
permitted lines from neutral and singly-ionized metals, including N\,{\sc i},
O\,{\sc i}, Mg\,{\sc i}, Mg\,{\sc ii}, Fe\,{\sc ii}, Ca\,{\sc ii} and Si\,{\sc ii}, and forbidden lines of
[O\,{\sc i}], [O\,{\sc ii}], [N\,{\sc ii}], [Fe\,{\sc ii}], [Ni\,{\sc ii}], [Ni\,{\sc iii}], [S\,{\sc iii}] and
[Ar\,{\sc iii}]. Weak lines of C\,{\sc i}, C\,{\sc ii} and [Fe\,{\sc iii}] are also observed, and
S\,{\sc i}, Ne\,{\sc i}, N\,{\sc ii}, Al\,{\sc ii}, [Cr\,{\sc ii}] and [Ar\,{\sc iv}] are tentatively
identified.  No compelling evidence for the presence of
higher-excitation lines over this wavelength range  is seen, with N\,{\sc iv}~\lam7114 and
He\,{\sc ii}~\lam1.012$\mu$m absent in the UVES spectra.

\begin{figure}
\begin{center}
\resizebox{\hsize}{!}{\includegraphics{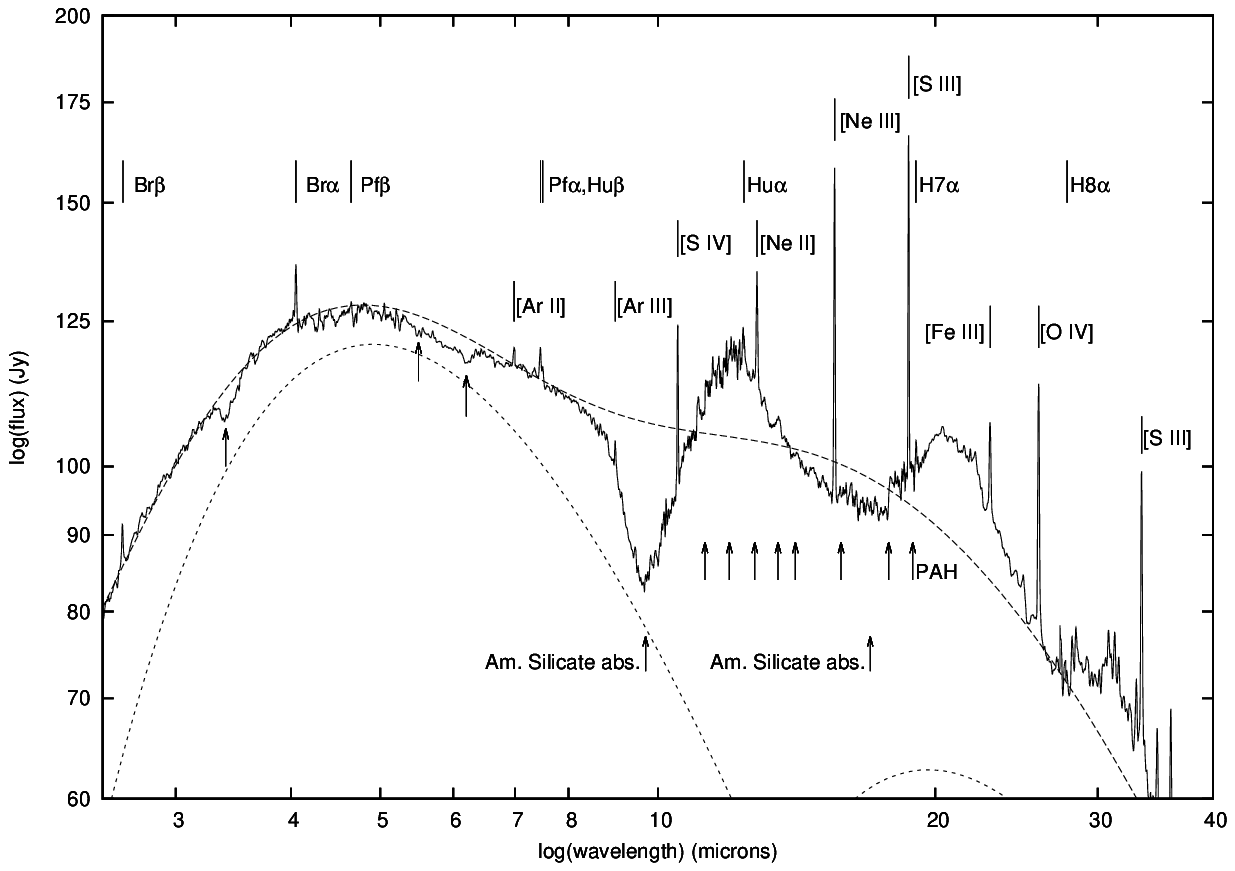}}
\caption{Log-scaled ISO-SWS spectrum of Wd1-9 covering
  2.4--20$\mu$m. Rest wavelengths of the principal emission lines and
  interstellar features are indicated, with the arrowed features $<$6microns 
being interstellar C2 and the PAH identifications from Boersma et al. (\cite{boersma}).
An illustrative  three-component black body fit comprising  a stellar source
  (25kK; not shown for clarity), hot ($\sim$1050K) and cool ($\sim$260K) 
dust (short dashed lines) and the summation of the contributions (long dashed line) is also shown.}
\label{fig:iso_spectrum}
\end{center}
\end{figure}

At longer wavelengths, the K-band spectrum of Wd1-9 reported by 
Mengel \& Taconni-Garman (\cite{mengel}) is 
essentially featureless, lacking the  CO bandhead emission that is seen in many other 
sgB[e] stars  (e.g. Liermann et al. \cite{liermann}); the $\sim6159$\Ang
TiO bandhead is likewise missing at shorter wavelengths. \object{Wd1-9} is 
also not identified as an emission line object in
He\,{\sc ii}(2.189$\mu$m)-\textit{continuum} or
He\,{\sc ii}(1.012$\mu$m)-\textit{continuum} interference filter photometry
(Crowther et al. \cite{crowther06}, Groh et al. \cite{groh06}). At  mid-IR 
wavelengths
Wd1-9 has a rich emission line spectrum superimposed   on a continuum dominated
by warm dust (Fig.~\ref{fig:iso_spectrum} and Sect. 4). Various hydrogen lines are present as are a number of forbidden lines from species such as 
[Ar\,{\sc ii}], [Ar\,{\sc iii}], [Ne\,{\sc ii}], [Ne\,{\sc iii}], [S\,{\sc iii}], [S\,{\sc iv}]  and [O\,{\sc iv}].

\begin{figure*}
\begin{center}
\resizebox{8.25cm}{!}{\includegraphics{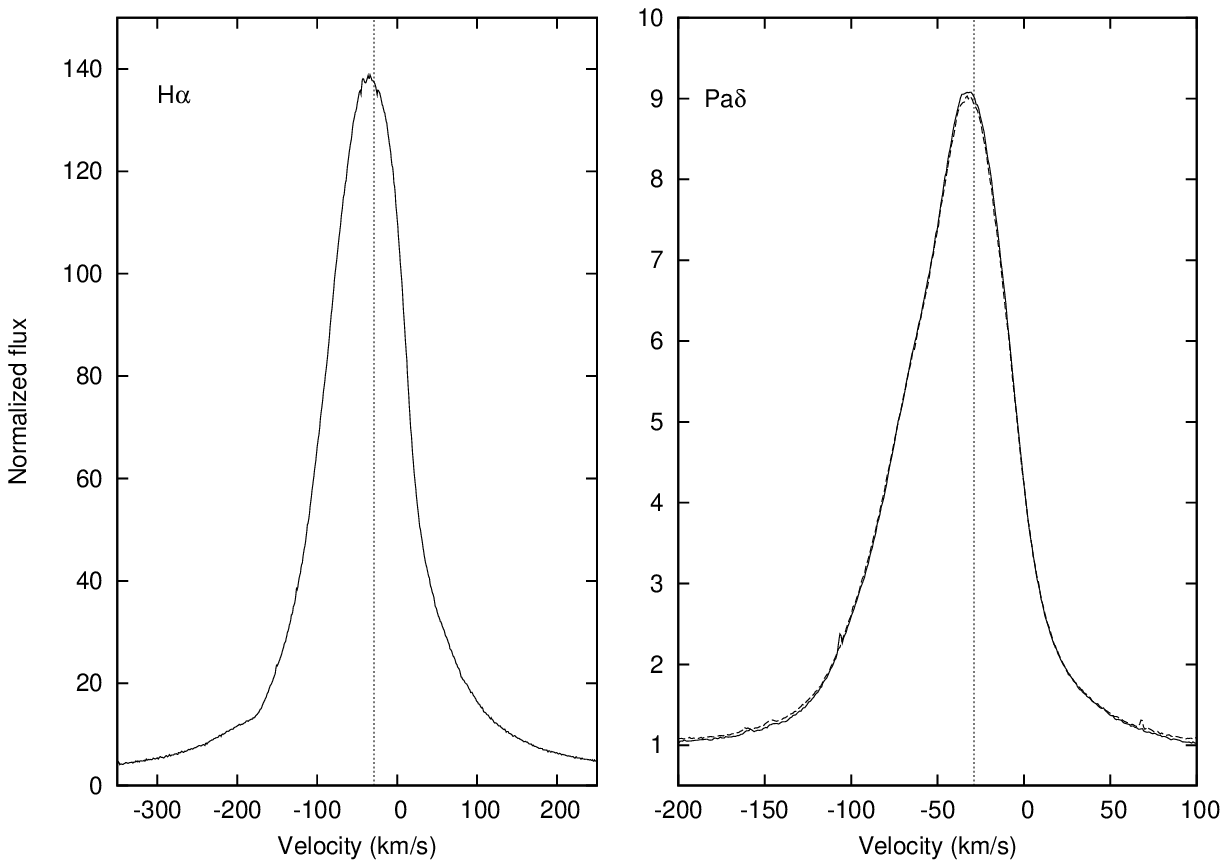}}\resizebox{8.25cm}{!}{\includegraphics{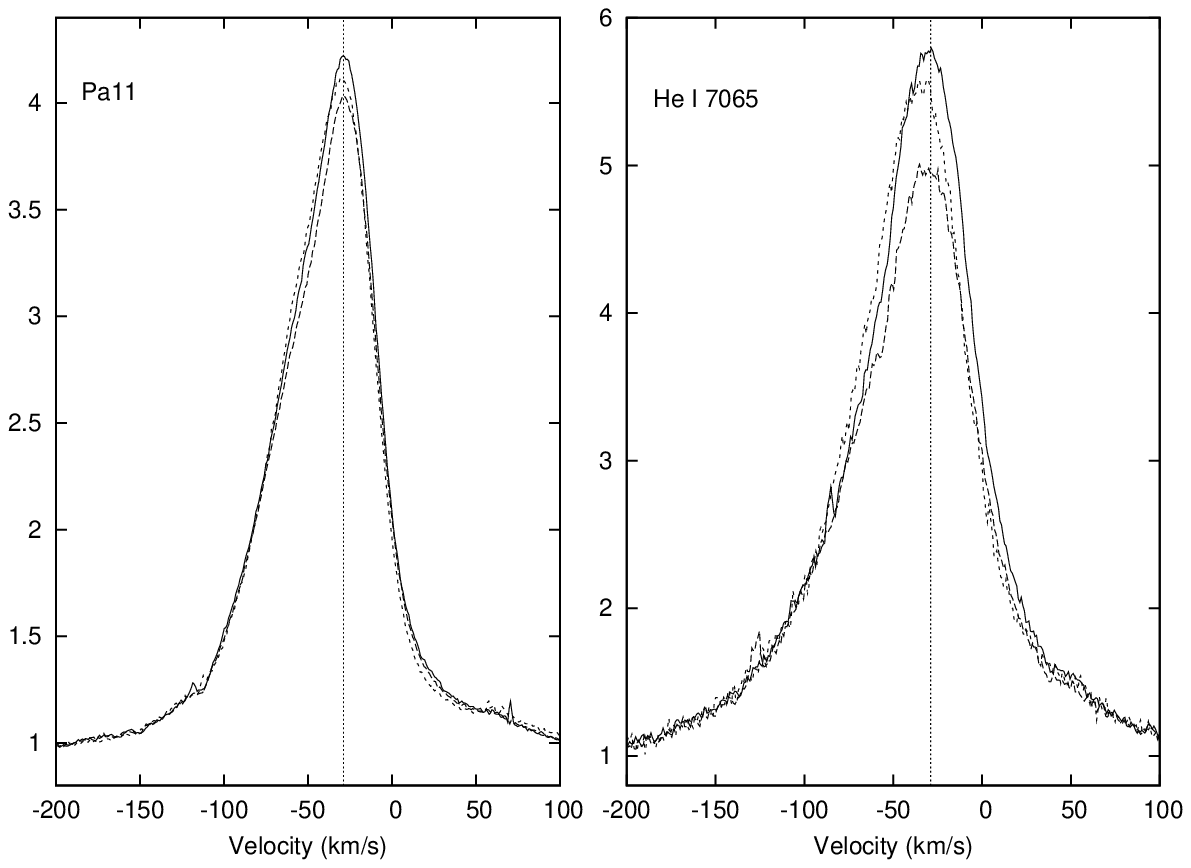}}
\resizebox{8.25cm}{!}{\includegraphics{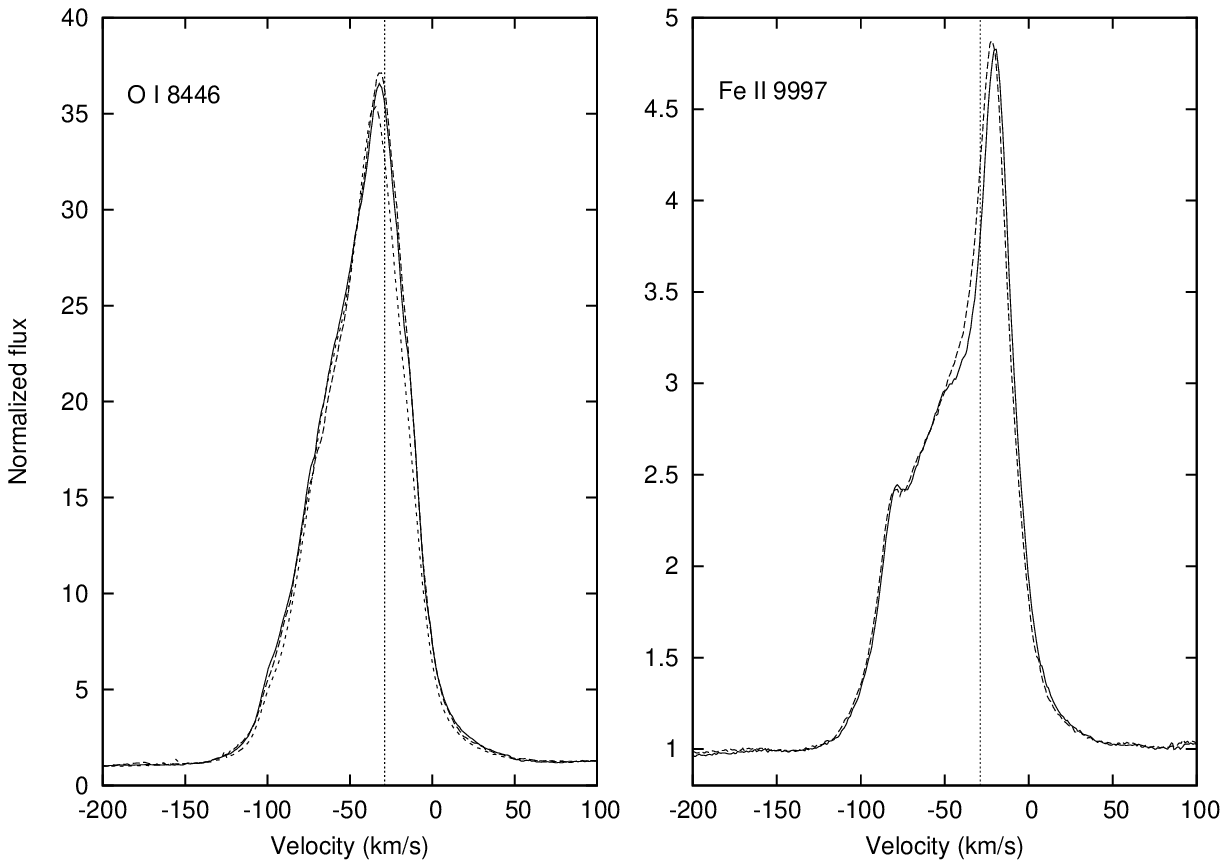}}\resizebox{8.25cm}{!}{\includegraphics{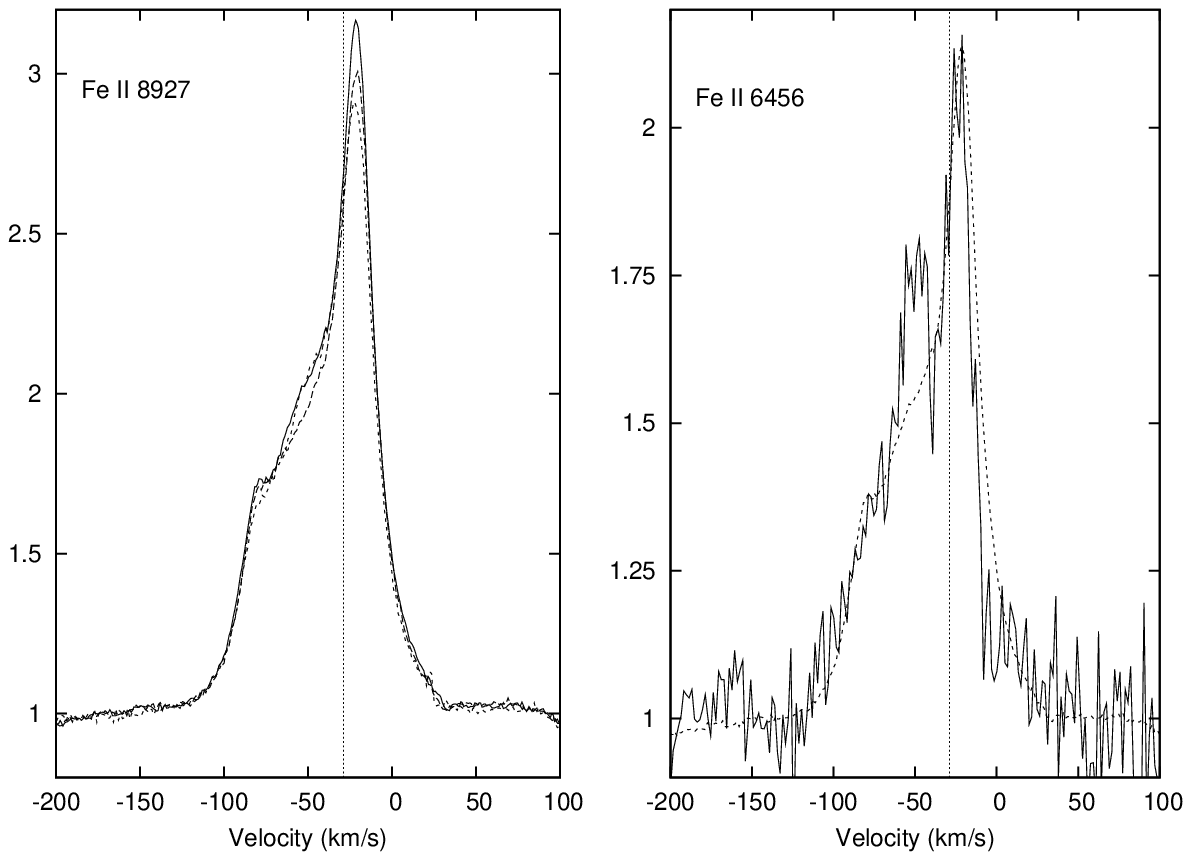}}
\resizebox{8.25cm}{!}{\includegraphics{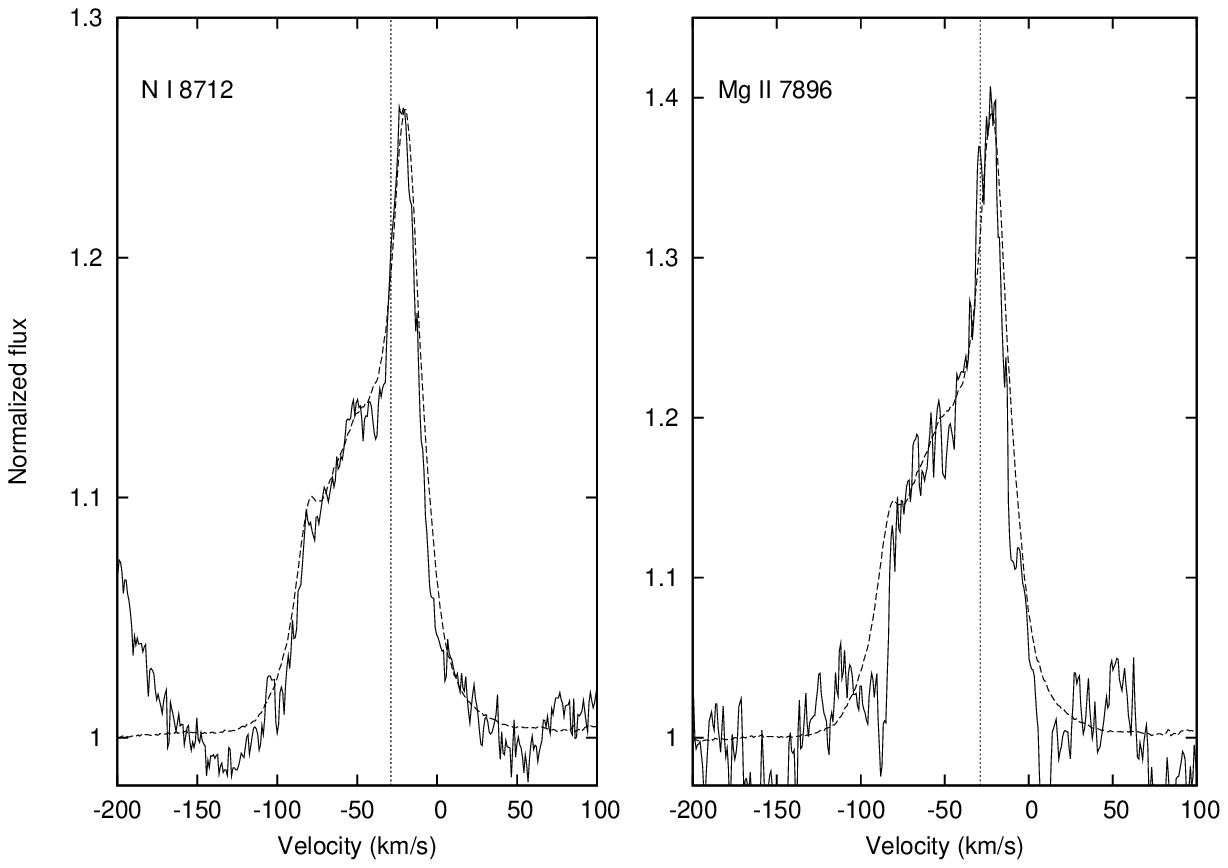}}\resizebox{8.25cm}{!}{\includegraphics{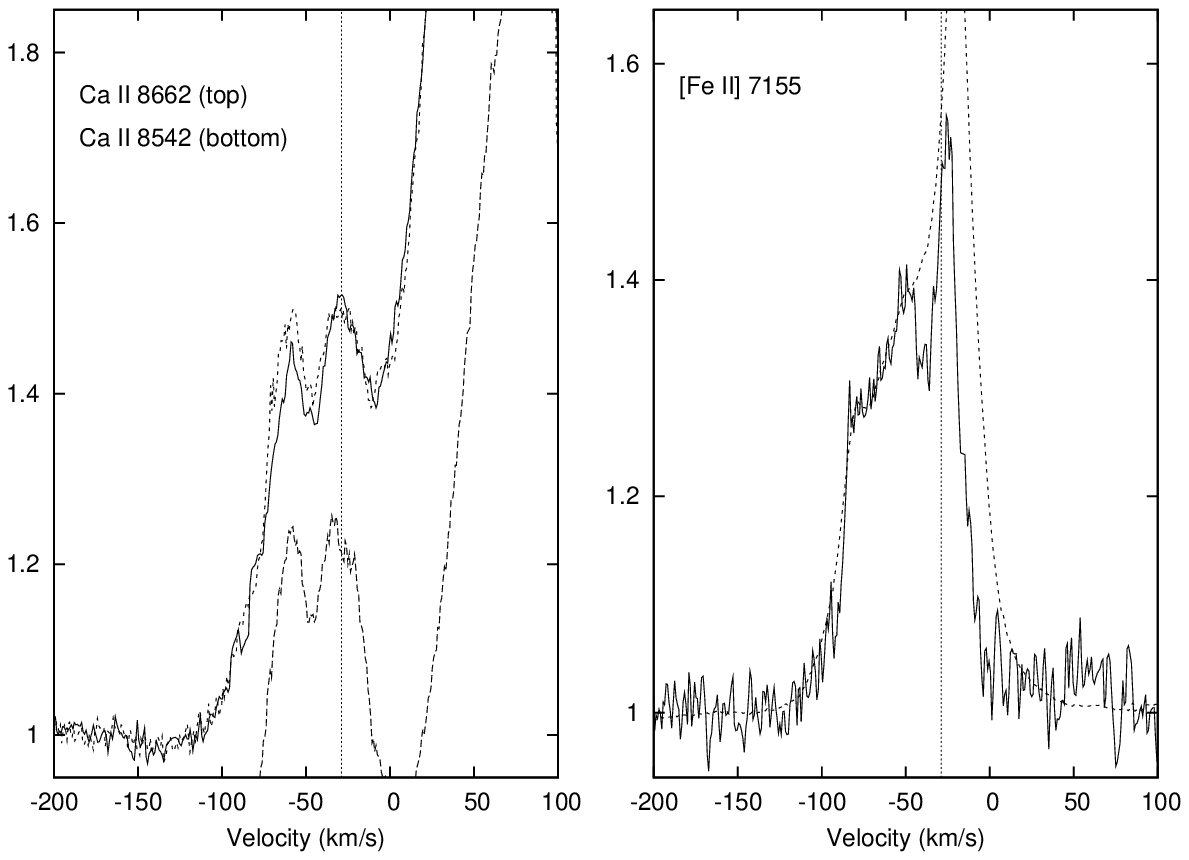}}
\resizebox{8.25cm}{!}{\includegraphics{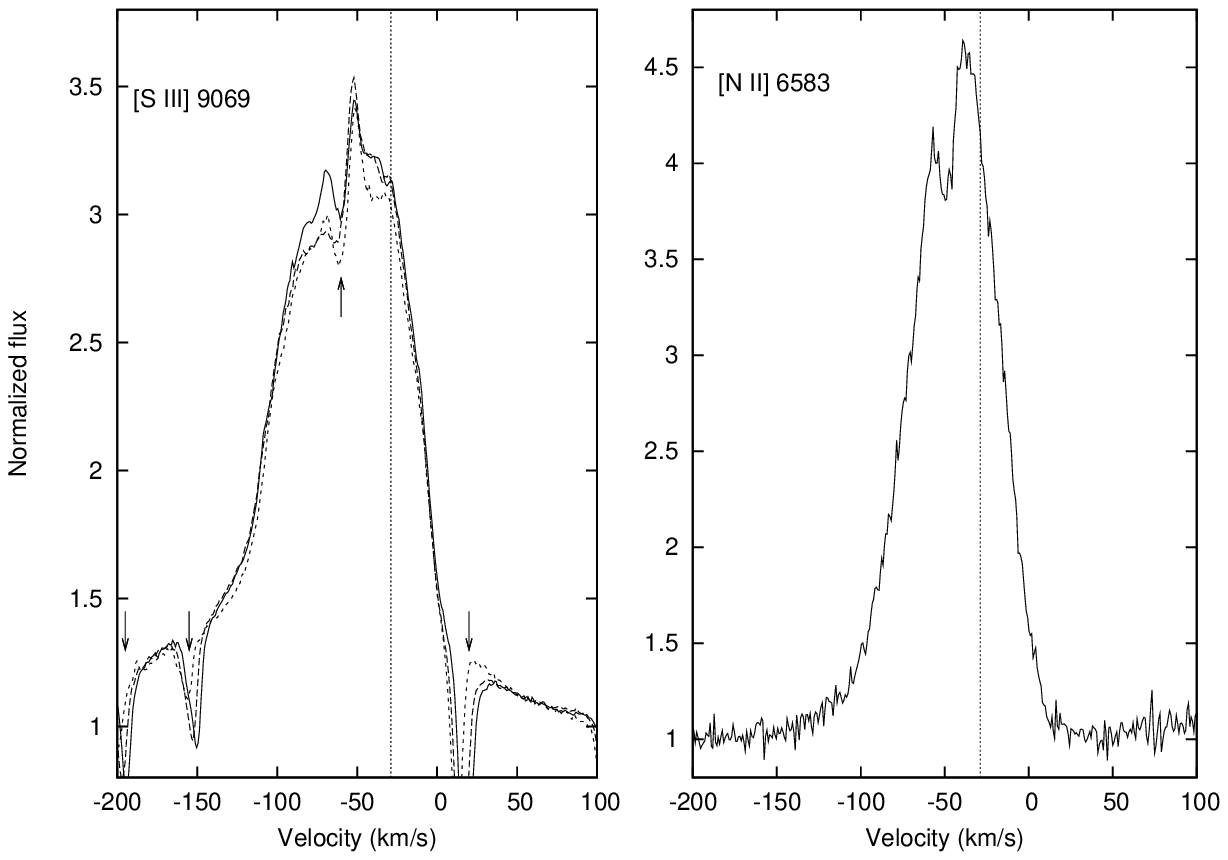}}\resizebox{8.25cm}{!}{\includegraphics{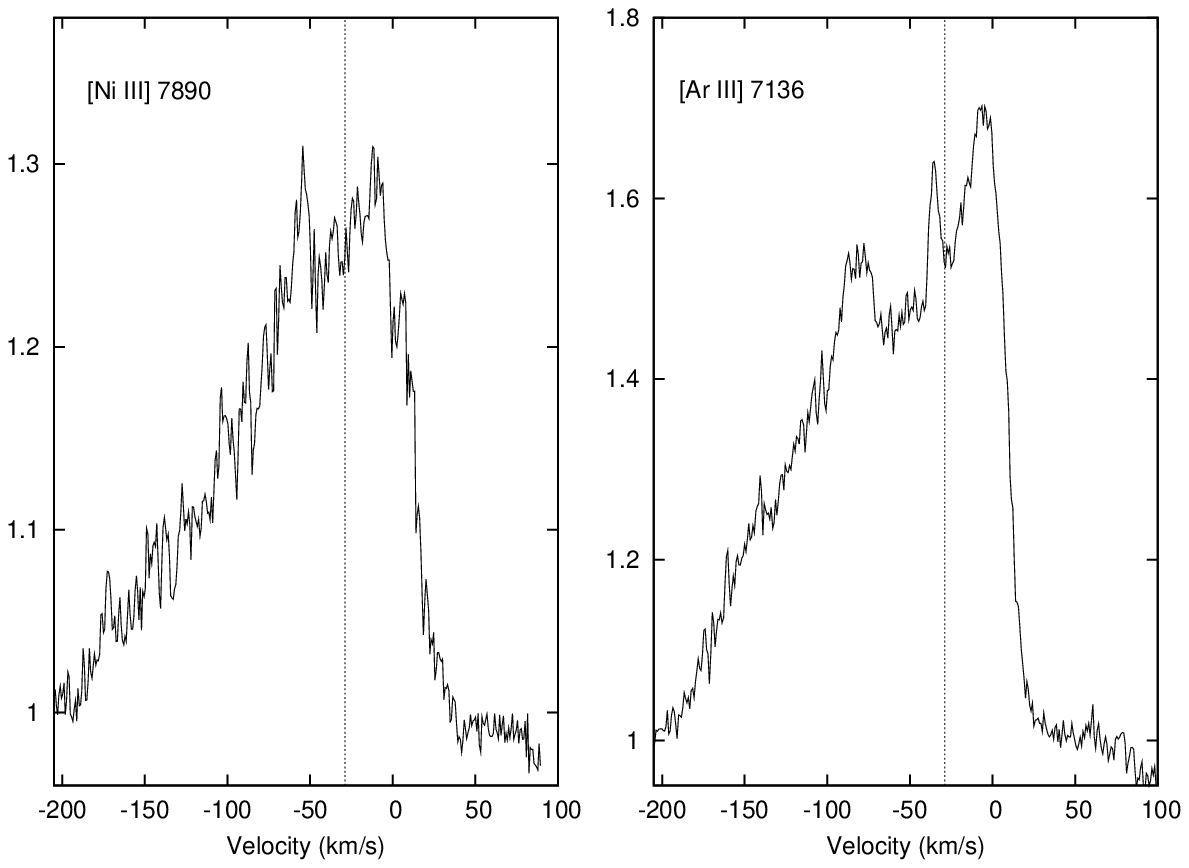}}
\caption{Strong permitted and forbidden emission lines in the spectrum
  of \object{Wd1-9}. Note that the H$\alpha$ (top left) has a different
  velocity scale to the remaining lines. In panels where multiple
  spectra are overplotted, solid line = 6/4/2011, dashed line =
  14/4/2011 and dotted line = 25/4/2011 respectively. For
  N\,{\sc i}~\lam8712, Mg\,{\sc ii}~\lam7896, Fe\,{\sc ii}~\lam6456 and [Fe\,{\sc ii}]~\lam7155 a
  scaled Fe\,{\sc ii}~\lam9997 line is overplotted for comparison, and in all
  panels the vertical line marks the radial velocity derived from the
  Pa9\dots16 lines. Arrows mark telluric features overlapping the
  [S\,{\sc iii}] line. }
\label{fig:lines}
\end{center}
\end{figure*}

\subsection{Hydrogen}

The dominant feature in the R-band spectrum of \object{Wd1-9} is an
extremely strong H$\alpha$ line (EW$=-640\pm40\text{\Ang}$) with a
relatively narrow core (FWHM$\sim$125\kms) and broad electron
scattering wings extending to $\pm$1500\kms, which appear asymmetric, with
excess emission in the blue wing  (Fig.~\ref{fig:lines}). 
The Paschen series is seen in emission to $n=35$, with the unblended
Pa9\dots16 lines yielding a radial velocity (RV) of $-28.9\pm0.8$kms$^{-1}$ 
that varies by less than 1\kms in the three epochs of UVES data.
Both H$\alpha$ and the Paschen series lines appear to be marginally
blueshifted in the 2004 FORS2 spectra, with a mean radial 
velocity
$-41\pm2$\kms derived for the Pa9\dots16 lines. Comparison with the
FORS2 spectrum from 2011 suggests that this is not just an
effect of the greatly improved resolution of the UVES data, although
the relatively low resolution of the earlier dataset preclude detailed
analysis.

At longer wavelengths, the ISO-SWS spectrum shows emission
from Br$\alpha + \beta$, Pf$\alpha + \beta$, Hu$\alpha + \beta$ and
the H7$\alpha$ and H8$\alpha$ transitions, although these lines are
much diminished by continuum emission from hot dust 
(Fig.~\ref{fig:iso_spectrum}).

\subsection{Helium}

\begin{figure}
\begin{center}
\resizebox{\hsize}{!}{\includegraphics{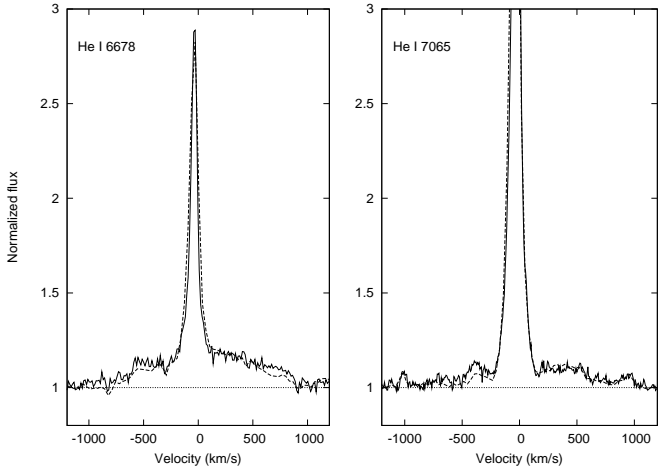}}
\caption{He\,{\sc i}~\lam\lam6678,7065 from UVES (solid line) and 2004 FORS2 (dashed
  line), showing the broad bases to the lines.}
\label{fig:He}
\end{center}
\end{figure}

Strong He\,{\sc i} emission is seen from the triplet He\,{\sc i}~\lamlam5876,7065
and singlet He\,{\sc i}~\lam6678 lines, while a weak He\,{\sc i}~\lam7281 line and
very weak I-band recombination lines from higher levels are also
observed. The He\,{\sc i} lines show the same blue-wing excess as the
H$\alpha$ and Pa$\delta$ lines, but also display broad emission bases
extending to at least $\pm$800\kms that are plotted in
Fig.~\ref{fig:He} (see also Clark et al. \cite{clark08}). The profile of the
base is broadly consistent between UVES and FORS2 spectra, although
the feature appears more pronounced at high velocities in the 2011
data; its presence in both datasets suggests it is both real and  not a
transient phenomenon. To the best of our knowledge these composite profiles are not
observed in any other sgB[e] star, although LHA 120-S 134 does demonstrate weak, broad 
He\,{\sc ii}$\lambda$4686 emission (Zickgraf et al. \cite{zick86}).

\subsection{Iron}
\label{sec:fe}

A large number of Fe\,{\sc ii} lines are visible in the spectrum of
\object{Wd1-9}, although no permitted lines of Fe\,{\sc i} or Fe\,{\sc iii} are
identified.  Fe\,{\sc ii}~\lam9997 line  is the most pronounced transition within
our spectral coverage, while Fe\,{\sc ii}~\lamlam8490,8927,9203 lines are
also strong. These lines originate from multiplets at $\sim$11eV,
implying strong Ly$\alpha$ pumping from metastable lower energy levels
(Hamann \& Simon \cite{hs86}, Sigut \& Pradhan \cite{sigut}). The Fe\,{\sc ii} 
lines display an asymmetric profile  (Fig.~\ref{fig:lines}),
with a strong, narrow red peak at $-20$\kms and a broad blue flank
with a sharp cutoff at $-80$\kms. Lines from multiplets 40 and 74
(e.g. Fe\,{\sc ii}~\lamlam6456,6515; upper energy levels $\sim$5--6eV) also
show a distinct second peak at $-50$\kms that is absent in the
Lyman-pumped lines.

A number of [Fe\,{\sc ii}] lines are present, with [Fe\,{\sc ii}]~\lam7155 being the strongest; this line also having a relatively high critical density of
 $\sim 10^8$cm$^{-3}$ (Zickgraf \cite{zick03}). The blue flank of
[Fe\,{\sc ii}]~\lam7155 is  identical  to the permitted Fe\,{\sc ii}
lines, while the red peak is less pronounced and significantly
narrower (Fig.~\ref{fig:lines}). 
Two weak [Fe\,{\sc iii}] lines from multiplet~8F are also identified in the optical spectra\footnote{Other [Fe\,{\sc iii}] lines that might be 
strong in \object{Wd1-9} lie too far bluewards to be accessible with the high reddening towards Wd1.}.
In contrast to [Fe\,{\sc ii}]~\lam7155, [Fe\,{\sc iii}]~\lam8838 
 demonstrates a wedge-shaped profile, with a
sharp red cut-off at $\sim$0~\kms and emission extending to almost
$-200$~\kms.

\subsection{CNO elements}
\label{sec:cno}

\object{Wd1-9} displays an extremely strong O\,{\sc i}~\lam8446 emission line
that has a similarly asymmetric profile to those of the strong Paschen-series
lines, and strong O\,{\sc i}~\lam7774 emission that displays a triple-peaked
profile as a result of the partial resolution of three component lines
separated by $\sim$3.5\Ang. The ratio \lam8446/\lam7774$\sim$20
implies that Ly$\beta$ fluorescence in a dense, predominantly neutral
transition zone between H\,{\sc i} and H\,{\sc ii} 
regions is responsible for the
O\,{\sc i}~\lam8446 emission (Grandi \cite{grandi80}, Kwan \cite{kwan84}),
 with the O\,{\sc i}~\lam7774
emission resulting from collisional excitation from the metastable
3s$^{5}$S$^{\circ}$ quintet ground state; a number of weaker O\,{\sc i} lines
from higher quintet levels are also observed.
In the 2011 data [O\,{\sc i}]~\lam6300 is weak and has a radial
velocity comparable to the other permitted lines, while [O\,{\sc i}]~\lam6363
is almost absent, but in the 2004 data both lines are much stronger,
with  asymmetric profiles and  redshifts of $\sim$40\kms compared to
the Paschen series (Fig.~\ref{fig:NaIOI}).

The  [O\,{\sc ii}]~\lamlam7319,7330 lines are present and 
display similar wedge-shaped profiles to [Fe\,{\sc iii}]~\lam8838.
 While heavy contamination by a telluric band
precludes detailed analysis, the feature appears broader than the [O\,{\sc i}] 
emission lines (Sect. 5). Finally
[O~IV] emission is observed in the mid-IR spectrum. This is  the 
highest-excitation feature in our spectra and 
 we discuss this further below.

N\,{\sc i} lines from multiplets~1,~2~and~8 are seen in the I-band spectrum,
while N\,{\sc ii} may also be present, but cannot be identified with certainty
due to blending. The N\,{\sc i} lines display near-identical profiles to
Fe\,{\sc ii}, with line intensity ratios of the N\,{\sc i} multiplet~1 lines
suggesting they are optically thick (Hamann \& Simon \cite{hs86}). Strong 
[N\,{\sc ii}]~\lamlam6548,6582 emission is also present, with double peaked 
line profiles that  appear to have 
weakened between 2004-11. Weak [C\,{\sc i}]~\lam6588
and [C\,{\sc i}]~\lamlam9094,9111 lines are also detected, with a profile
similar to N\,{\sc i}, while broad, weak C\,{\sc ii}~\lamlam7231,7236 lines 
are also present, although heavily blended with a telluric feature.

It appears likely that the N\,{\sc i} and C\,{\sc i} lines form in a cool
dense environment.  The N\,{\sc i} multiplet 1 and 2 and C\,{\sc i} multiplet
3 lines originate from excited states that are linked to the ground state
by strong UV resonance lines and in a high density environment 
($\geq10^{10}$cm$^{-3}$) the resonance lines become very optically thick (Hamann \& Simon \cite{hs88}), allowing the population of excited states to become 
large; collisional excitation can then populate the upper energy levels of the observed emission lines. In contrast, the [N\,{\sc ii}] lines 
have low critical densities ($<10^5$cm$^{-3}$) and must therefore trace a physically distinct lower-density region.

\subsection{Sodium and Calcium}
\label{sec:CaMgNa}

\begin{figure}
\begin{center}
\resizebox{\hsize}{!}{\includegraphics{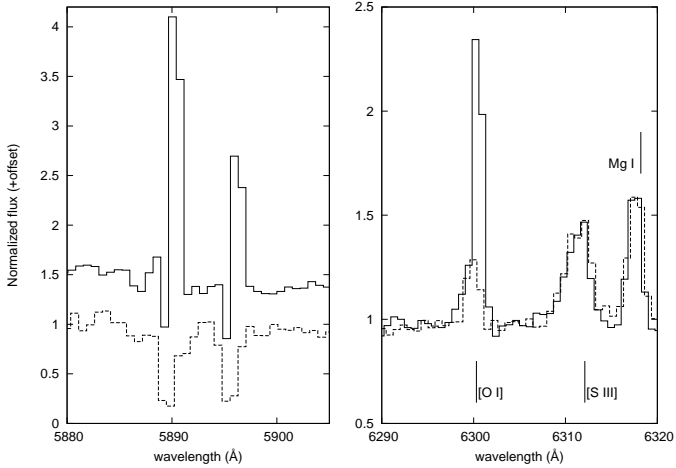}}
\caption{Comparison FORS2 spectra from 2004 (solid line) and 2011
  (dashed line), for Na\,{\sc i}~\lamlam5890,5896 (\textit{left panel}) and
  [O\,{\sc i}]~\lam6300 (\textit{right panel}). }
\label{fig:NaIOI}
\end{center}
\end{figure}

Na\,{\sc i} and Ca\,{\sc ii} emission lines all arise from low-excitation 
states, implying the presence of a cool region shielded from the ionizing flux
responsible for the Lyman-pumped lines. In the 2004 FORS2 data strong
Na\,{\sc i}~\lamlam5980,5986 emission lines were present, redshifted by
$\sim$40\kms compared to the other permitted lines.   Strong
interstellar absorption bluewards of the emission line centre gives
the lines apparent P~Cygni profiles at this time, but the emission 
component is
completely absent in the corresponding 2011 data, with only the 
interstellar components remaining (see
Figure~\ref{fig:NaIOI}; mirroring the behaviour of [O\,{\sc i}]).
 Ca\,{\sc ii}~\lamlam8498,8542,8662 lines lie on the
blue wing of adjacent Paschen-series lines and cannot be separated in
the 2004 FORS2 spectrum, but are clearly resolved by UVES, showing
double-peaked profiles separated by 26~\kms (Fig. 4); these lines are the only
permitted lines to show clear double peaks. The corresponding
[Ca\,{\sc ii}]~\lamlam7291,7324 lines that link the lower levels of the
near-IR permitted lines to the Ca\,{\sc ii} ground state are not apparent,
implying $n_e$$>$10$^7$cm$^{-3}$ in the Ca\,{\sc ii} line forming region.

\subsection{Forbidden lines}
\label{sec:forbidden}

In addition to the  forbidden line emission described above, strong 
[S\,{\sc iii}]~\lam\lam9069,9531 lines were observed, along with 
[Ni\,{\sc ii}]~\lamlam7378,7412, [Ni\,{\sc iii}]~\lamlam7890,8500,
[Ar\,{\sc iii}]~\lamlam7136,7751, and [S\,{\sc iii}]~\lam6312.  Very weak
[S\,{\sc ii}]~\lam1.287,1.320$\mu$m and [Cr\,{\sc ii}]~\lam8125 lines 
are tentatively detected in the UVES
spectrum;  [Ar\,{\sc iv}]~\lam7237 may be also be present, but is
blended with a telluric band.

The observed lines span a range of ionziation energies and critical
densities, indicating that they do not trace a single
line-formation region, and this is reflected in their profiles. 
  [Ni\,{\sc ii}]~\lamlam7378,7412 lines are weak but  clearly
double peaked with a separation $\sim$25\kms, resembling Ca\,{\sc ii} in this 
respect. In contrast the higher excitation
 [Ni\,{\sc iii}]~\lam7890,8500 and [Ar\,{\sc iii}]~\lamlam7136,7751 lines 
resemble the wedge-shaped profiles of  [O\,{\sc ii}]~\lamlam7319,7330 and 
[Fe\,{\sc iii}]~\lam8838. [Ar\,{\sc iii}] displays a complex
triple-peaked profile with narrow secondary components at $-41$ and
$-87$\kms superimposed on  broad emission (Fig.~\ref{fig:lines}) while
[Ni\,{\sc iii}] only displays a single narrow component at $-54$\kms. 
Finally  [S\,{\sc iii}]~\lam9069 is
very strong, and appears generally similar to [N\,{\sc ii}] but with a
broader profile, complex peak structure
 and wings extending to at least $-200$/$+100$\kms.

The ISO-SWS spectrum displays a number of strong fine-structure 
lines that include [Ar\,{\sc ii}], [Ar\,{\sc iii}], [Ne\,{\sc ii}], 
[Ne\,{\sc iii}], [S\,{\sc iii}], [S\,{\sc iv}] and the aforementioned  
[O\,{\sc iv}]. Emission in [Ne\,{\sc iii}], [S\,{\sc iv}] and [O\,{\sc iv}] 
suggest rather higher temperatures, with 41eV required to ionize Ne$^+$ and
 54eV required to ionize O$^{++}$. These lines would require an ionising 
source with temperature $\sim$60--80kK, but this hard to reconcile with the 
lack of He\,{\sc ii} or N\,{\sc iv} emission. However [O\,{\sc iv}] emission 
also arises in post-shock gas with $n_e$$\sim$10$^3$cm$^{-3}$ and $T$$\sim$50kK
(Lutz et al. \cite{lutz98}) and  we consider this a more likely origin for the
highest-excitation forbidden lines visible in the ISO-SWS spectra.

\subsection{Variability}

Despite the presence of rapid aperiodic photometric variability (Bonanos 
\cite{bonanos07}) and the inference from radio observations  of a 
significantly higher mass loss rate in the past (Dougherty et al. 
\cite{dougherty}), one of the remarkable features of the current dataset is 
the lack of spectroscopic variability across the course of the 30 years of observations. This stands in contrast  to 
other sgB[e] stars such as LHA 115-S 18 and  luminous evolved stars in general (e.g. Clark et al. \cite{clark10}, 
\cite{clark13}); Wd1-9 
shows no evidence of the long-term changes in gross spectral morphology characteristic of
LBVs and cool hypergiants,  although the sgB[e] star LHA 115-S 65 likewise demonstrated
no variability for 3 decades before the sudden appearance of CO bandhead emission (Kraus et al. \cite{kraus10}, Oksala et al.
\cite{oksala}).

He\,{\sc i}~\lam7065  is the
\textit{only} line in  our spectra to show  the  significant short-term
variability in strength - but not in  radial velocity -  that 
is characteristic of emission lines in other luminous early-type stars as a result of their 
 asymmetric and unstable winds\footnote{He\,{\sc i}\lamlam5876,6678 are only
  observed in the single UVES spectrum using dichoric \#1, so cannot
  be checked for variability.}. Apart from this, only the low excitation
 Na\,{\sc i}\lamlam5890,5896, [O\,{\sc i}]~\lam\lam6300,6363 and 
[S\,{\sc ii}]~\lamlam6716,6730 and the higher excitation [N\,{\sc ii}]\lam\lam6548, 6582 
lines have been found to vary, being  substantially weaker post-2004.
Finally, we reitterate  that with the single exception  of the 2004 FORS2 spectrum we
 were unable to determine any RV changes in the system from analysis of the 
 Pa9\dots16 lines; a conclusion also reached by Cottaar et al. 
(\cite{cottaar})  from analysis of the strong emission lines of their  spectra.
We therefore conclude that Wd1-9 does {\em not} appear to demonstrate the reflex RV motion 
expected of a binary system in these transitions, although we have insufficient data to determine whether
the strength of the He\,{\sc i}~\lam7065 transition is periodically modulated.

\section{The dusty circumstellar environment of Wd1-9}

\label{sec:ir}
The ISO-SWS spectrum of \object{Wd1-9} (Fig. 3) shows pronounced similarities to those of other sgB[e] stars 
such as R66 and R126, suggesting comparable  circumstellar envelope compositions and geometries. 
Specifically, the comparatively flat mid-IR spectra of these stars are  indicative of multi-temperature, 
silicate-rich  circumstellar  dust and stand in contrast to the much `bluer' spectra of normal  mass-losing (red) 
supergiants (Kastner et al. \cite{kastner}, \cite{kastner10}). 
 An {\em illustrative} three-component black body
fit to the spectrum of Wd1-9 is plotted in Fig. 3, comprising
 a $\sim$25kK stellar source\footnote{The fit is only weakly dependant  on the choice of stellar temperature and a  range corresponding to early- to mid-B supergiants can be accomodated.} and
two dust components with temperatures $\sim$1050K and $\sim$260K,  and 
provides a close match to the overall IR spectrum. However,  we caution that a series of
silicate absorption and emission features in the 8--35$\mu$m range
(cf. Chiar \& Tielens \cite{chiar}) leads to uncertainty in the placement of the continuum
level, particularly  for the cool dust component. 

The profile of the 9.7$\mu$m
absorption component is very similar to that present in the ISO-SWS spectra of  the
heavily interstellar reddened WC8-9 Wolf-Rayet stars \object{WR98a} and \object{WR112}
($A_V{\sim}9.9$ and 10.2 respectively; van der Hucht et al. \cite{vdh}). The comparable reddening of Wd1 
cluster members 
(mean $A_V{\sim}10.4$; Negueruela et al. \cite{negueruela10}) leads us to suppose that the main 
contribution to this feature in Wd1-9 is also interstellar in origin, although we cannot exclude an additional
 circumstellar component. A  number of weak hydrocarbon absorption features are present at
3.4$\mu$m, 5.5$\mu$m and 6.2$\mu$m; these features are also likely to be 
interstellar, as other indicators  of a carbon-rich dust chemistry
within the \object{Wd1-9} circumstellar environment are absent.

\begin{figure}
\begin{center}
\resizebox{\hsize}{!}{\includegraphics{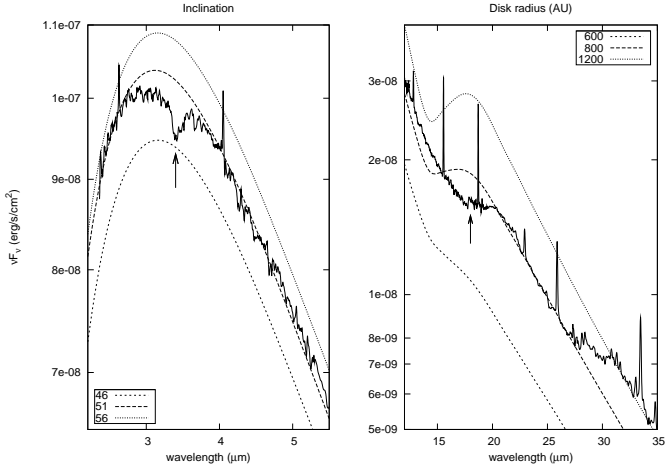}}
\caption{Comparison of models for varying input parameters. ({\it Left
    Panel}) Effect of varying inclination. ({\it Right Panel}) Effect
  of varying the radius of the disk. Absorption features
  due to carbon (3.4$\mu$m) and amorphous silicates
  ($\sim$18$\mu$m) are marked.}
\label{fig:iso_dullemond}
\end{center}
\end{figure}

No discontinuity is apparent at the $\sim$29$\mu$m transition between
ISO-SWS bands~3 and~4, implying that the cool dust component is not
greatly extended\footnote{The ISO-SWS Band~3 aperture is 27$\times$14"
  while the Band~4 aperture is 33$\times$20"; an extended cool
  component results in a discontinuity due to the inclusion of
  $\ge$30$\mu$m emission from a region larger than the Band~3 aperture
  (Lamers et al. \cite{lamers96}).} and mid-IR imaging  also shows
that dust is located close to \object{Wd1-9} (Clark et al. \cite{clark98}). Given this, we
make use of the simple semianalytic irradiated disk model of Dullemond et al. 
(\cite{dullemond}) to examine the geometry and distribution of dust around \object{Wd1-9}
 in an identical manner to Kastner et al. (\cite{kastner}) in their analysis of 
 the sgB[e] star \object{R126}. 

The model is based on a flared disk with an inner `hole', with three disk
regions: a hot, `puffed up' inner rim that is directly exposed to the
stellar flux, a cool intermediate region that is in the shadow of the
rim and an outer flared region that is again illuminated by the
central star. The model does not include the multi-dimensional
radiative transfer required to fully model a dusty object like
\object{Wd1-9}, nor the effect of interstellar absorption on the emergent spectrum
 and hence considerable caution must be employed in interpreting the results of this analysis in light of 
the preceding discussion.
Nevertheless such an approach is useful in constraining the broad properties of
the system and permits a direct comparison to be made to the disc properties of  R126 
(Kastner et al. \cite{kastner}).

 We fix the distance and luminosity of \object{Wd1-9} at $5kpc$
from Negueruela et al. (\cite{negueruela10}), adopt T$_*$=25kK for the central source  and, following Kastner
 et al. (\cite{kastner}), fix the  hot dust temperature at 1050K from the  black-body fit to the 
short wavelength region of the spectrum.
We find that the overall spectral energy distribution is reproduced
well by the model, with inner and outer radii of $\sim60AU$ and $\sim800AU$ respectively,
 dust mass$\sim$4$\times$10$^{-4}M_{\odot}$ and inclination $\sim$50 degrees (Fig.~\ref{fig:iso_dullemond}). 
The visibility of
the hot inner wall of the disk represents the primary constraint on
inclination, with both face-on and edge-on configurations showing
weaker near-IR emission and a greater far-IR excess than intermediate
viewing angles (see also Dullemond et al. \cite{dullemond} on the effects of varying
inclination). Similarly, the far-IR energy distribution depends on the
extent of the cool outer reaches of the disk, with more compact
configurations leading to lower emission beyond 20$\mu$m due to the
reduced emitting volume of cold dust; inconsistent with our observations.
Subject to contamination by the assumed interstellar absorption features, the 
broad emission complexes at $\sim 10 \mu$m and  $\sim 20\mu$m (Fig. 3)
are reproduced by optically thin emission from the surface layers of the disc, 
as found by Kastner et al. (\cite{kastner}; their Fig. 3) for R126.

Mindful of the limitations of our model, it is instructive to  compare the results for  Wd1-9 to those obtained by
Kastner et al. (\cite{kastner}) for R126. As expected, given the similarities in the spectra, the dusty discs in 
both systems appear broadly comparable. However, that associated with Wd1-9 appears less 
massive than that of 
R126 in terms of dust mass ($4\times10^{-4} M_{\odot}$ versus $3\times10^{-3}M_{\odot}$) and
spatial extent ($r_{out} \sim 800AU$ versus $\sim 2500AU$), although it extends closer to the star ($R_{in}\sim 60AU$ versus $\sim120AU$) 
and hosts correspondingly hotter dust ($\sim 1050$K versus $\sim 800$K). Nevertheless 
in both cases this is significantly lower than the expected dust sublimation
temperature ($\sim1500$K), implying a central cavity. Given the isolation of R126, it is 
tempting to attribute the lower mass and size of the Wd1-9 disc to the harsher 
environment in which it resides due to  the diffuse UV and X-ray radiation fields of Wd1. In this regard the commetary 
nebulae associated with the RSGs within Wd1 are also suggestive of ablation of their 
outer atmospheres by the action of the cluster wind/radiation field (Dougherty et al. 
\cite{dougherty}).

\section{Wd1-9 as a sgB[e] star}

Synthesising the results of the preceding sections, we may construct a 
picture of a highly structured and complex circumstellar environment 
associated with  Wd1-9:

\begin{itemize}
\item A cool, dense ($\sim 5000$K, $>10^8$cm$^{-3}$;
Hamann \& Simon \cite{hs88}) region which is  shielded from an ionising flux -
and hence hydrogen neutral - in which the Ca\,{\sc ii}, C\,{\sc i}, N\,{\sc i}
and Mg\, {\sc i} form. The collisionally-excited Fe\,{\sc ii} lines with upper 
energy levels of $\sim 5-7$eV would also arise here. 
\item A hotter, dense transitional zone between the H\,{\sc i} and 
H\,{\sc ii} regions ($\sim 8000$K, $10^9-10^{10}$cm$^{-3}$) where hydrogen is 
partially ionised ($\tau(\rm{H}\alpha)>10^3$; Grandi \cite{grandi80}) in which 
Ly$\alpha$ pumping of Fe\,{\sc ii} and Ly$\beta$ pumping of 
O\,{\sc i}~\lam8448 and Mg\,{\sc ii} proceeds.
\item A hot H\,{\sc ii} region in which the H and He lines form.
\item A relatively cool ($> 5000$K), low-density and velocity region where 
narrow singly-ionised forbidden lines of [Fe\,{\sc ii}], [O\,{\sc ii}] and [N\,{\sc ii}] form. The significant differences in
critical densities for these lines (e.g. [Fe\,{\sc ii}]\lam 7155 versus 
[N\,{\sc ii}]\lam 6583) suggests that this region is itself structured.
\item A hot, low density ($\sim 30,000$K, $<10^6$cm$^{-3}$; Hamann \& Simon 
\cite{hs88}) region where relatively broad doubly ionised forbidden lines of 
[Ar\,{\sc iii}] and [S\,{\sc iii}] form.
\item A high temperature, diffuse, post-shock  region ($\sim 50,000$K, 
$n_e \sim 10^3$cm$^{-3}$: Lutz et al. \cite{lutz98}) responsible for the  highest 
excitation fine structure lines such as [O\,{\sc iv}]. 
\item A cool, massive dusty torus of radius  $\sim 60$ to $\sim 800AU$ and 
mass $\sim 4\times10^{-2}M_{\odot}$ assuming a canonical gas to dust ratio 
of 100.
\end{itemize}

and, including the results of previous studies, we may also add:  

\begin{itemize}
\item A partially optically thick steady-state outflow with an $r^{-2}$ density distribution and 
(unclumped) mass loss rate of $9.2^{+0.4}_{-0.5}\times 10^{-5}M_{\odot}$yr$^{-1}$ (Dougherty et al.  \cite{dougherty}).
\item A compact H\,{\sc ii} region of inner radius $4000AU$ 
indicative of enhanced mass loss 
( $33^{+12}_{-5}\times 10^{-5}M_{\odot}$yr$^{-1}$) which terminated $\sim200$yr ago 
(Dougherty et al.  \cite{dougherty}).
\item A region of very high temperature gas
($kT \sim 2.3^{+0.5}_{-0.3}$keV; Clark et al. \cite{clark08}).
\end{itemize}

The diversity observed in the line profiles also emphasises the complexity of the circumstellar environment (Fig. 4). On this basis, while we do 
not
attempt to fully reconstruct the kinematics and geometry of the Wd1-9 system from these  data, they do offer several 
additional insights. Firstly, no shell emission lines
are present in the spectra, implying that Wd1-9 is not seen edge-on; consistent with the mid-IR modeling. No P Cygni profiles indicative 
of pure outflow are present in the hydrogen or helium lines, although the broad bases to He\,{\sc i}\lam\lam 5876,  
6678 and 7065 are suggestive of a high velocity outflow. The remaining line profiles are harder to interpret. The asymmetries seen in the line 
profiles are common to B[e] stars and  classical Be stars. In the latter objects they are attributed to the presence of 
global one armed oscillations in the gaseous circumstellar discs (e.g. Okazaki \cite{okazaki}); in  sgB[e] stars one must also account for the presence of stellar or disc winds 
and self-absorption by the  circumstellar disc/torus (Oudmaijer et al. \cite{oud98b}, Zickgraf \cite{zick03}). 

Nevertheless, the presence of double 
peaked emission lines in e.g. the Ca\,{\sc ii} and [Ni\,{\sc ii}] lines is suggestive
of an origin in a quasi-Keplerian disc or torus (e.g. Kraus et al. \cite{kraus10},  Aret et al. \cite{aret}). Likewise the higher excitation permitted
lines  appear to be broader than the low excitation forbidden lines, as seen in other sgB[e] stars (Zickgraf \cite{zick03}) and also 
 consistent with this hypothesis. 
This is also apparent via comparison of the [O\,{\sc i}] and [O\,{\sc ii}] lines, with the former narrower than the latter, as predicted for such a scenario  by Kraus et al. (\cite{kraus12}). 

However, the asymmetric profiles of  [O\,{\sc ii}]\lam 7319 and higher excitation species such as 
[Ni\,{\sc iii}]\lam7890 and [Ar\,{\sc iii}]\lam7136 also appear to 
show a departure from emission from  an axisymmetric quasi-keplerian disc, due to the presence of an additional broad emission component. 
Somewhat surprisingly,  a simple correspondence between line width and excitation does not appear to 
be present: e.g.  [N\,{\sc ii}]\lam6583 (excitation potential of 14.53eV)  is
  narrower than [Fe\,{\sc ii}]\lam7155 (excitation potential of 7.9eV). 
Indeed, in conjunction with excitation requirements and critical densities, comparison of the profiles of these lines 
(and also [Ar\,{\sc iii}]\lam7136 versus [S\,{\sc iii}]\lam9069) suggest that the physically distinct regions itemised  above are themselves structured.

\object{Wd1-9} therefore conforms well to the
canonical picture of B[e] stars, with a dense, low-ionization
equatorial region, possibly in quasi-Keplerian rotation,  in which the low excitation permitted and forbidden lines
form (Zickgraf et al. \cite{zick85}, Lamers et al. \cite{lamers}, Zickgraf et al. \cite{zick03}, Kraus et al. \cite{kraus10}, Aret et al.
\cite{aret}). Higher-excitation lines appear to demonstrate an additional contribution 
which one might speculate arises in a  radial outflow (stellar- or disc-wind?), while 
we may associate some   lines (such as O\,{\sc i}~\lam8446, [Fe\,{\sc ii}]~\lam7155 and the
strong Ly$\alpha$-pumped Fe\,{\sc ii} transitions) with a transitional zone on the surface of the disk,
where densities are high but a significant ionizing flux is present. Finally, Wd1-9 appears
encircled by a detatched massive dusty disc extending out to large radii.

\section{The evolutionary state of Wd1-9}

Presumably the result of the complete veiling of the `central engine' of Wd1-9 by circumstellar
material, the  absence of any photospheric features and apparent lack of RV
changes indicative of orbital motion in Wd1-9  complicates its physical interpretation from  optical 
spectroscopy. 
Consequently, the primary constraints on both the
nature and evolutionary state of \object{Wd1-9} come from radio (Dougherty et al. \cite{dougherty}) and X-ray observations
(Skinner et al. \cite{skinner}, Clark et al. \cite{clark08}). The former implies
a recent epoch of eruptive mass loss at rates significantly in excess of 
$\sim$10$^{-4}M_{\odot}$yr$^{-1}$ that subsequently decreased by a factor of $\sim4$           a few hundred years ago
(Dougherty et al. \cite{dougherty}). Such extreme mass loss rates are found only in extreme cool hypergiants
(Lobel et al. \cite{lobel03}, Humphreys et al. \cite{hump05}), LBV outbursts
(Davidson \& Humphreys \cite{davidson}, Groh et al. \cite{groh10}) and the fast
phases of Roche-lobe overflow in interacting binary stars
(Langer et al. \cite{langer03}, Petrovic et al. \cite{petrovic}), being $\sim$2 orders of magnitude greater than 
expected for the late-O/early-B supergiants that characterise Wd1 at this time (Negueruela et al. \cite{negueruela10}; 
Crowther et al. \cite{bsg}).

The X-ray observations are potentially a more powerful diagnostic. The spectrum is best fitted by a thermal plasma model with $kT\sim2.3$keV, and yields
a integrated luminosity of $\sim10^{33}$ergs$^{-1}$; a combination of temperature and luminosity that is
 clearly inconsistent with emission from a single massive star (Clark et al. \cite{clark08}). 
Therefore, despite the absence of cyclic photometric or spectroscopic (RV) variability
indicative  of binarity, the X-ray luminosity and
plasma temperature appear to require a colliding-wind system or a
high-mass X-ray binary (HMXB).

X-ray detections of sgB[e] stars are rare  (Bartlett et al. in prep.). Two {\em bona fide} HMXB with sgB[e] star 
primaries are known - CI Cam (e.g. Clark et al. \cite{clark00}, Hynes
et al. \cite{hynes}) and IGR J16318-4848 (Filliatre \& Chaty \cite{filliatre}) - while a dusty circumstellar 
environment has also been   inferred for a further two - GX301-2 (Moon et al. \cite{moon}) and SS433 (Fuchs et al. 
\cite{fuchs}). While the X-ray luminosity of Wd1-9 is consistent with the first two systems (but is significantly
 fainter than the latter two) its thermal X-ray spectrum  is very different from the hard, power law spectra of these HMXBs  (Clark et al. \cite{clark08}, Bartlett et al. \cite{bartlett})\footnote{Two additional X-ray detections of 
 Magellanic Cloud sgB[e] stars have been made (S18 and S134; Clark et al. \cite{clark13}, Bartlett et al. in 
prep.)but low X-ray count rates  preclude a determination  of the nature of their  putative binary companions.}.

In contrast to the HMXBs discussed above, Wd1-9 bears considerable resemblance in terms of both X-ray spectrum and 
luminosity
to known compact, colliding wind binary systems within Wd1, such as the 
 ($P_{\mathrm{orb}}\sim$)3-day WN7+O binary \object{WR-B}, the 4-day O+O
binary \object{Wd1-30a} and the 7-day WN7b+OB binary
\object{WR-A} (Bonanos \cite{bonanos07}, Clark et al. \cite{clark08}, Ritchie et al. \cite{ritchie13}).
Indeed RV observations (Ritchie et al. \cite{ritchie09a}, \cite{ritchie13}) show that \textit{all} 
OB supergiants within Wd1 with similarly hard  X-ray spectra to \object{Wd1-9} are binaries
with periods less than ten days. A similarly short-period O6-7~V+O7--8~V progenitor system 
($M_{\mathrm{initial}}\sim 40+35M_{\odot}$) for Wd1-9 would 
therefore fit  both its current observational properties and the wider evolutionary
context of Wd1 (cf. the eclipsing WNLh+O binary Wd1-13; Ritchie et al. \cite{ritchie10}).

If \object{Wd1-9} is the short-period binary suggested by the X-ray data, it is not expected to
undergo either an LBV or RSG phase. It will instead experience highly non-conservative
mass transfer as the primary evolves off the main sequence and
overflows its Roche lobe (Petrovic et al. \cite{petrovic}),  leading to the ejection of the
hydrogen envelope and direct formation of a very short period WR+O
binary (cf. \object{WR-B}; Bonanos \cite{bonanos07}) without the system
undergoing a loop redwards across the Hertzsprung-Russell diagram (Clark et al. 
\cite{clark11a}). Theoretical models of massive binary evolution by Wellstein \& Langer (\cite{wellstein}) and
Petrovic et al. (\cite{petrovic}) indicate brief phases of \textit{fast} Roche-lobe overflow mass loss at
(time-averaged) rates $\gtrsim 10^{-4}M_{\odot}$yr$^{-1}$; directly comparable to that
inferred for Wd1-9 (Dougherty et al. \cite{dougherty}).

Following the  fast phase of Roche-lobe overflow, mass loss rates decrease
until either the surface hydrogen
abundance of the primary drops below $X_s$$\sim$0.4 and a Wolf-Rayet
wind develops (Petrovic et al. \cite{petrovic}) or core hydrogen burning ends and a second
phase of \textit{fast} transfer takes place as the mass donor
commences shell burning (Langer et al. \cite{langer03}). Depending on how far mass
transfer has progressed, both Wolf-Rayets and O9--B0 (super)giants
therefore represent plausible current primaries  for Wd1-9. A WC star is incompatible with
the carbon-poor dust chemistry indicated by the infra-red
observations, while an early WN star would display strong He\,{\sc ii} and
N\,{\sc iv} emission that is not apparent in our spectra. In contrast, the
very weak photospheric lines of the interacting O+O binary
\object{Wd1-30a} suggest that such a configuration would be completely
undetectable against the rich emission line
spectrum of Wd1-9, while weak He\,{\sc ii} emission from a WN9--11h+O system would also likely be
undetectable. Such a system would
ultimately emerge from the sgB[e] phase as an WNL+O binary, appearing
similar to \object{Wd1-13} (Ritchie et al. \cite{ritchie10}), WR-B (Bonanos \cite{bonanos07}) and eventually, after 
further mass loss,
\object{WR-F} (Clark et al \cite{clark11a}). We therefore emphasise that  {\em if} Wd1-9 is indeed a 
compact interacting binary, viable progenitor {\em and} descendent
systems are present  within  Wd1.

\subsection{Corrobarative evidence}

\subsubsection{Comparator systems}

The suggestion that the B[e] phenomenon (in part) arises due to binary-driven mass loss is not new\footnote{See for example 
Langer \& Heger (\cite{langer98}),  Sheikina (\cite{sheikina}), Zickgraf (\cite{zick03}), Kastner et al. (\cite{kastner},\cite{kastner10}), 
Miroshnichenko 
(\cite{miro07}), Kraus et al. (\cite{skraus}, \cite{kraus12}).} and a number of lines of observational
 evidence/comparator systems support this assertion. Analogues of the detatched dusty  toroid surrounding the central
binary of Wd1-9 (Sect. 4) have been  resolved interferometrically around the sgB[e] stars  HD327083 
(Wheelwright et al. \cite{wheelwright}) and   MWC300 (Wang et al.  \cite{wang}), with the former of considerable interest since 
it comprises a binary {\em within} the torus.   Although  less massive and with larger orbital
separation than expected  for 
Wd1-9, GG Car and VFTS~698 are both spectroscopic sgB[e] binaries (Gosset et al. \cite{gosset}, Dunstall et al. \cite{dunstall}), 
with the former associated with a circumbinary molecular disc/torus (Kraus et al. \cite{kraus12}).
While not formally classified as a sgB[e] star, HDE 326823 is a likely $\sim 6.123$day binary (Richardson et al. \cite{richardson}), comprising a proto-WN8 primary
 (Marcolino et a. \cite{marcolino}) and a more massive
secondary  hidden by an accretion disc, with both stars in turn surrounded by a circumbinary disc with a dusty component 
(Ardila et al. \cite{ardila}). Similarly
\object{RY~Scuti} is an 11-day eclipsing binary consisting of a Roche lobe-filling
$8M_{\odot}$ O9.7 supergiant and  $30M_{\odot}$ companion  associated with a circumbinary disc ($r \sim1AU$) nested within
 a toroidal nebula  ($r \sim10^3AU$),
 delineated by sequential  radio and IR emission from ionised gas and warm dust 
(Smith et al. \cite{smith02}, Grundstrom  et al. \cite{grundstrom}, Smith et al. \cite{smith11}). 

Given the similarity in evolutionary state of the latter two objects to that proposed for Wd1-9 a more detailed
comparison is of interest.
While a quantitative determination of the geometry of the circumstellar/-binary envelope of  
 HDE 326823 has yet to be achieved, RY~Scuti appears to  differ from Wd1-9 in several aspects. Specifically, the 
 inner bound of the (dust free) circumbinary disc in RY~Scuti is over an order 
of a magnitude  smaller than we infer for Wd1-9 (Sect. 4, Grundstrom et al. \cite{grundstrom});  
conversely   the outer toroidal nebula of 
RY~Scuti is significantly larger than the Wd1-9 dusty toroid. Given that the 
dust in the toroidal nebula of  RY~Scuti is already cooler than that found for Wd-9 and other sgB[e] stars (Kastner et al. 
\cite{kastner}), it  
will {\em not} evolve into a similar  sgB[e] configuration unless fresh 
dust production occurs in the inner circumbinary disc. A further difference is that the radio emission 
in Wd1-9 is 
associated with ongoing mass loss from the system, while in RY~Scuti it arises from the detatched nebula.
One might speculate that these discrepancies arise from differences in the  mass-transfer pathways - and the stage of  
progression along them - experienced by both systems;   the $\sim25+15M_{\odot}$ initial 
configuration suggested for   \object{RY~Scuti} implies a lower
 mass than that of  Wd1-9, (Smith et al. \cite{smith11} and  Sect. 6.0) and consequently 
that mass transfer in RY Scuti may have proceded via a more conservative mode.

A better match to Wd1-9 may be provided by MWC349A. Binarity has been suggested by several studies  (Hofmann et
 al.  \cite{hofmann}, Gvaramadze \& Menten \cite{gvara}) and the circumstellar/-binary geometry appears similar 
to Wd1-9, comprising a warm, dusty Keplerian torus ($r\sim100AU$, $T\sim600-1000$K; Danchi  et al. \cite{danchi}, Kraemer et al. 
\cite{kraemer}, Weintroub et al. 
\cite{weintroub}) that  is associated with a bipolar disc-wind visible at radio wavelengths  (White \& 
Becker \cite{white}). Both Wd1-9 and 
MWC349A also share similar optical and  mid-IR spectra (Kraemer et al. 
\cite{kraemer}, Hamann \& Simon \cite{hs88}) and while  the latter lacks the very high excitation lines and hard 
X-ray emission of the former, this might be the result of absorption by the dusty torus in MWC349A, which is viewed edge on and hence 
obscures the central engine. 

So to conclude; while we are able to identify a number of systems with features comparable to those of Wd1-9 we are not able to provide 
an exact match. However it appears that binary driven mass transfer can yield  the defining observational features of the sgB[e] 
phenomenon  and given the rapidity of this phase and the sensitivity of the process to initial conditions (e.g. component masses and 
orbital period; Petrovic et al. \cite{petrovic}) and indeed inclination, this is possibly unsurprising.  

\subsubsection{The association of sgB[e] stars with massive stellar clusters}

A clear prediction of the hypothesis that a subset of sgB[e] stars form via binary interaction is that since
the mass transfer phase is rapid ($\sim 10^4$yr; Petrovic et al. \cite{petrovic}) such stars should be rare. 
Is this the case? An obvious methodology to assess this is to search for 
{\em bona fide} sgB[e] stars  in young massive clusters, since this permits an accurate determination of 
stellar luminosity and also enables a direct comparison to the populations of other classes of  massive stars\footnote{We note that in doing so we are implicitly associating the duration of rapid case A mass transfer with that of 
the sgB[e] phase. We are not aware of any studies regarding the lifetime of the dusty (Keplerian?) discs 
of sgB[e] stars, although following the analysis of Bik et al. (\cite{bik}) regarding the photoevaporation of 
proto-stellar discs around OB stars and allowing for the additional effects of X-ray irradiation and a cluster
UV radiation field one might expect them to be short-lived ($\lesssim 10^5$yr).}.
The evolved stellar population of Westerlund 1 is certainly consistent with such an hypothesis; Wd1-9 is the only 
sgB[e] star amongst $\geq$100 OB supergiants and large 
numbers   of other supposedly short lived objects such as hot and cool hypergiants and Wolf-Rayet  
(Negueruela et al. \cite{negueruela10}, Ritchie et al. \cite{ritchie13}). 

Spectroscopic surveys of massive stars are available for $\sim$68 young ($\leq 20$Myr) stellar aggregates containing $\sim600$ massive post-MS stars, enabling us to build upon this approach (Sect.  A.1.).
While evolved stars of all known spectral types are present within these clusters,
 only Wd1 and  Cyg OB2 are found  to host sgB[e] stars, while they are also the least frequently encountered 
post-main sequence stars in  purely numerical terms (Table A.1). Trivially, these conclusions hold over any age subsets of the 
 sample (for example RSG dominated clusters at $>10$Myr).
Turning to the Magellanic clouds, Massey et al. (\cite{massey00}) associate sgB[e] stars with
two of the 19 MC clusters they study. LHA 120-S 134 appears a {\em bona fide} example and its spectral and X-ray 
properties  point to it being a massive binary (Bartlett et al. in prep.). However Bonanos et al. (\cite{bonanos09})
question the identification of [L72] LH85-10 as a sgB[e] star, prefering a classical Be star classification on the basis of luminosity and  shape of the IR continuum. Recently, four B[e] stars have been identified within the 30 Doradus star forming complex; VFTS 039, 698, 822 
and 1003 (Evans et al. \cite{evans}, in prep.). However, since
 star formation is still occurring within this non-coeval complex,  distinguishing 
between sgB[e] and HerbigB[e] classifications for these stars is 
difficult,  although, if CO bandhead emission is present, enrichment of $^{13}$CO relative to $^{12}$CO may in principle be used as
a discriminant (Kraus \cite{kraus09}, Liermann et al. \cite{liermann}).
 In any event, and  as with Wd~1, with at most 4 candidates drawn from over 700  hot luminous stars  
(Doran et al. \cite{doran}) the stellar population of 30 Dor also  argues for the intrinsic rarity  of sgB[e] stars, as indeed do the dearth 
of 
clusters in both the Galaxy and Magellanic Clouds that host them.

If the B[e] phenomenon is an evolutionary state that \textit{all} supergiants pass
through (with progenitor masses spanning $\sim12M_{\odot}$ for the RSG clusters  to $\sim100M_{\odot}$ for the Arches cluster)
then the problem appears acute; the sgB[e] phase must be extremely brief, such that the chances of
observation are low even in clusters containing a rich population of high-mass stars. 
In contrast, if the phenomenon is (in part) associated with
rapid, binary-mediated mass transfer, then the absence of sgB[e] stars
in these clusters is easier to understand: while still an intrinsically rapid process, 
 in the youngest  clusters only a limited  subset of stars will be both  compact binaries and sufficiently evolved to have 
 reached the onset of mass transfer. While it might be supposed that the standard (Kroupa) initial mass 
function should lead to greater numbers of  
interacting binaries in the older, RSG dominated clusters, we suspect that the tendency towards
conservative mass transfer in lower-mass systems
(cf. Wellstein \& Langer \cite{wellstein}) may preclude the extensive mass loss
required to yield the sgB[e] phenomenon. Likewise, the huge physical extent of
the RSGs that characterise these clusters  means that such stars must be either isolated or in such wide
binary systems that the secondary does not influence the evolution of
the RSG primary\footnote{If sgB[e] stars were exclusively post-RSG objects one would 
anticipate their  absence from  clusters  of this age, where stars are 
expected to end their lives as RSGs. 
However, the 
presence of low luminosity  sgB[e] stars
 (e.g. log($L_{\mathrm{bol}}$/L$_{\ast}) <5.0$; Gummersbach et al. \cite{gummersbach}) implies that 
such comparatively low mass  ($<20M_{\odot}$) stars may indeed encounter
  this phase.}.

\section{Conclusions}
\label{sec:concl}

A detailed synthesis of radio, infra-red, optical and X-ray observations of
\object{Wd1-9} are consistent with a hot, binary stellar source
surrounded by a massive dusty disk viewed at moderate inclination. Suprisingly 
little evidence for long- or short-term spectroscopic variability is present
at this time, despite apparent low-level aperiodic  photometric modulation and historical
indications of a long term evolution in the mass loss rate. Taken as a whole, 
the multiwavelength observational properties of Wd1-9 are consistent with those of other members of the 
rather heterogeneous class of  sgB[e] stars, with the
{\em caveat} that the presence of high excitation [Si\,{\sc iv}] and [O\,{\sc iv}] appears unprecedented.
Nevertheless, the most likely explanation for their occurence - that Wd1-9 is a massive interacting 
binary and they are associated with shocked gas resulting from the wind/wind collision zone - is not without precendent (cf.
LHA 115-S 18; Clark et al. \cite{clark13}).

Due to the apparent lack of variability, no orbital period can be 
identified for \object{Wd1-9}; we suspect this is due to the almost 
complete veiling of the central binary by the circumstellar/-binary envelope. Nevertheless the X-ray properties of Wd1-9 provide a 
compelling case for binarity; {\em all} stars  with comparable X-ray properties in Wd1 are confirmed
 O+O or WR+O colliding-wind binaries, with periods less than ten
days (Ritchie et al.  \cite{ritchie10b}, \cite{ritchie13}). Of these,  the 4-day interacting O+O
binary \object{W30a} and the 7-day WN7+O binary \object{WR-A} have essentially identical X-ray
spectra to that of \object{Wd1-9} in terms of both morphology and
luminosity. Crucially, if Wd1-9 is to be interpreted in terms of an ongoing or recently completed episode of binary-driven
mass-transfer we are able to find both viable progenitor and descendant systems within Wd1 (Sect. 6).

While we cannot  exclude the possibility that Wd1-9 is a quiescent HMXB, a 
more natural explanation for its observational properties is that it is a short-period O+O or WNVL+O binary that has
recently undergone extreme mass loss during Roche-lobe overflow
(Petrovic \cite{petrovic}), and is now buried within a circumbinary disk.
Specifically, the extreme historical mass loss
rate inferred from radio observations (Dougherty et al. \cite{dougherty})
is of a magnitude comparable with time averaged theoretical predictions for fast Roche-lobe overflow in 
Case A (core H-burning) binaries (Wellstein \& Langer \cite{wellstein}, Petrovic et al. \cite{petrovic}). Likewise, our best-fit 
model to the spectral energy distribution of Wd1-9 is consistent with a detatched dusty torus surrounding the central engine,
 replicating the findings of Kastner et al. (\cite{kastner}), who interpreted this geometry in terms 
of a circumbinary Keplerian disc resulting from binary driven mass loss. 

Critically, recent spectroscopic and interferometric observations have identified comparator short-period 
binary systems surrounded by detatched (dusty) discs (Sect. 6.1.1), while the apparent rarity of sgB[e] stars (Sect. 6.1.2) is 
also qualitatively consistent with the expected rapidity of mass-transfer in compact binaries ($\sim10^4$yr in its most extreme 
phase; Petrovic et al. \cite{petrovic}). 
We therefore suspect that  binary interaction provides both a significant  formation channel for evolved sgB[e] stars and  a natural 
explanation for the large range of luminosities exhibited by such stars 
(e.g. Gummersbach et al. \cite{gummersbach}, Miroshnichenko \cite{miro07} Aret et al. \cite{aret}).
 Indeed the apparently unbiased luminosity distribution of sgB[e] stars
  argues against their evolution 
 from  a mass-restricted subset of massive stars, in contrast to the 
occurence of phases such as the cool hypergiants  and WN7-9ha stars.  The diversity of observational  properties of 
sgB[e] stars would then  result from a combination of the initial binary configuration - and hence precise mass transfer route followed -
and time since interaction.

In any event, Wd1-9 provides us with a unique opportunity to observe such a binary apparently `caught in the act', with the 
rich stellar population of its host cluster allowing us to place the system in a proper evolutionary context. As such it is likely
 to become a cornerstone system for quantitatively constraining the physical process of rapid binary-driven mass-transfer in 
massive evolved stars.

\begin{acknowledgements}

We thank Michiel Cottaar for providing spectra of Wd1-9 obtained in
August 2009 and July 2010. JSC gratefully acknowledges the support of
an RCUK fellowship.  IN has been funded by grant AYA2010-21697-C05-05  from the Spanish 
Ministerio de
Ciencia e Innovaci\'on (MICINN).

\end{acknowledgements}

\appendix

\section{Summary of the evolved stellar content of Galactic young massive clusters}

 In order to determine the relative frequency of occurence of sgB[e] stars in comparison to other post-MS sub-types
we  constructed a {\em representative} census of well-studied young ($\leq$20Myr), massive
 Galactic clusters. To accomplish  this we employed the source lists of
Massey et al. (\cite{massey95}, \cite{massey}), Eggenberger et al. (\cite{eggenberger}),  Crowther et al. 
(\cite{crowther06}), Evans et al. (\cite{evans}), Davies et al. (\cite{davies12b}) and Chen\'{e} et al. (\cite{chene}) to undertake a literature survey\footnote{Given the time that has elapsed, the properties of many of the clusters studied by these authors have been revised. This is apparent for e.g. Eggenberger et al. (\cite{eggenberger}), where new determinations of cluster ages have placed a number of clusters 
outside our 20Myr cut-off (cf. NGC 3766, NGC 3590, NGC 5281, NGC 6664  and Trumpler 18), while a combination of (i) revisions in the  classification of individual stars  and  (ii) the inclusion of stars of spectral-type A in their blue supergiant category 
explain both the discrepancies in number counts of red and bue supergiants between the works and the omission of additional clusters from our census (e.g. NGC 2384, NGC 6530, Trumpler 37 and Collinder 121, all of which now appear not to host supergiants).}, supplemented with the 3 
Galactic Centre clusters and a number of aggregates associated with Giant H\,{\sc ii} regions\footnote{M17, W3, W31, W42, W51, 
G333.1-0.4, NGC3576,   Sh2-152, S255, and RCW34}; as expected the latter possessed no unambiguous post-MS stars and so are not considered further. This resulted in a total of $\sim68$ stellar clusters and/or associations
and where possible we present both cluster mass and age in Table A.1, along with associated uncertainties if quoted
 in the literature.

For simplicity we group stars into eight broad  spectral classifications in Table A.1, based on our current 
understanding of the evolutionary sequence of post-MS stars; a more detailed breakdown is superfluous, given 
 our simple aim of qualitatively determining the rarity, or otherwise, of sgB[e] stars. This results in a total of $\gtrsim600$ stars.

Inevitably both the cluster census and the populations of individual clusters summarised in Table A.1  are likely to be incomplete and subject to 
observational biases that are difficult to quantify given the diverse detection strategies employed by the various authors; hence we consider the values quoted to be {\em lower limits}. Specifically:

\begin{itemize}
\item The location of  RSGC1-6 within a wider star forming region similarly dominated by RSGs complicates assessment of cluster extent/membership. Moreover, interstellar extinction mandates their study in the far-red/near-IR, biasing any surveys against the detection  of blue supergiants and WRs, noting that population synthesis arguements would suggest broadly comparably numbers of blue and red supergiants in 10-20Myr-old clusters  (e.g. Davies et al. \cite{davies09}).
\item More generally, the lack of systematic narrow-band imaging surveys for obscured clusters will hamper identification of intrinsically faint WRs, while the weak winds of lower-luminosity OB supergiants will not yield 
 pronounced emission lines and hence will prevent their detection via such an observational 
strategy. We suspect that the second limitation   is particularly  problematic for the  Galactic Centre proper and Quintuplet clusters, and is further compounded by the low S/N of available spectra for the latter aggregate, to the extent that in many cases we are unable  to confidently 
determine accurate spectral types and/or luminosity classes for individual members
(Figer et al. \cite{figer}, Liermann et al. \cite{liermann09}).
\item Any single-epoch observational survey will be biased against identifying LBVs; one might anticipate that both late-B supergiants and YHGs could be cool-phase LBVs. 
\item The qualitative classification criteria used to distinguish between yellow super- and hyper-giants.
\end{itemize}

Nevertheless we suspect that sgB[e] stars will be amongst those least affected by incompleteness; their strong 
emission lines, intrinsic brightness and characteristic near- to mid-IR excess (e.g. Bonanos et al. \cite{bonanos09}) 
render them easily identifiable. While we recognise the overlap in spectral morphologies of sgB[e] stars and cool-phase 
LBVs in the near-IR window (cf. S18 and GG Car versus AG Car; Morris et al. \cite{morris}) comparison of both near- and mid-IR 
colours allows for their separation (Zickgraf et al. \cite{zick86}, Bonanos et al. \cite{bonanos09}). Moreover, they may be distinguished  from the spectroscopically similar but less evolved Herbig AeBe stars on the  basis of their intrinsic luminosities.

Bearing the above {\em caveats} in mind, the results of this compilation appear broadly as expected (cf Davies et al. \cite{davies09}), with younger clusters being dominated by the apparently massive core 
H-burning  early-mid O supergiants and WNLha stars and older clusters hosting increasing numbers of red supergiants. Both WN and WC subtypes are present in young ($<10$Myr) clusters; it is curently not possible to determine whether their absence in older clusters
is a result of observational bias due to interstellar reddening and/or stellar evolution. Both LBVs/BHGs/WN9-11h stars and YHGs are present in small numbers- as anticipated due to the apparently short durations of these phases - in clusters spanning a comparatively wide range of ages
($\sim2-12$~Myr). This appears to support the predictions  of e.g. Groh et al. (\cite{groh13})  that both high and low mass stars pass through such evolutionary phases (the former always remaining in the blue region of the HR diagram and the latter on the bluewards track of a red loop).

Only two
 instances  of clusters hosting confirmed sgB[e] stars are found - Wd1 and the Cyg OB2 association (MWC349A). Two further stars are of interest. Firstly Martayan et al. (\cite{martayan}) identify  one member of NGC 6611 - the emission line binary W503 - as demonstrating an IR excess, which the authors attribute to the presence of a mass-transfer accretion disc. Further observations of this object to better constrain its nature would  be of considerable value. Secondly,  a detatched equatorial ring is
associated with the $\sim$4~Myr old B1.5 Ia star Sher25, located in the outskirts of the young ($\sim$1~Myr) cluster NGC3603 (Brandner et al.  
\cite{brandner}). Although the  current stellar spectrum lacks the forbidden lines characteristic of sgB[e] stars (Smartt et 
al. \cite{smartt}), Smith et al. (\cite{smith07}) suggest that Sher 25 might represent the evolved descendant of such a star.

Finally we strongly caution against employing these data in order to {\em quantitatively} determine
 the progenitor masses of evolved stars. In several cases the ages of clusters have been determined solely from the presence 
of particular spectral types; therefore  deriving  progenitor masses for such stars from the age of the cluster introduces
 a circularity into the argument. Moreover in many of the remaining cases ages have been inferred from isochrone fitting to 
sparse datasets, while in the near-IR window favoured for the study of  obscured clusters isochrones are essentially vertical, 
presenting an additional difficulty. 
Finally, it is increasingly clear that binarity plays a major role in post-MS 
evolution (e.g. Clark et al. \cite{clark11a}); given that systematic RV surveys have only been attempted for a handful 
of clusters any detailed analysis of the data presented here would appear premature.

\longtab{1}{
\begin{longtable}{lccccccccccr}
\caption{Stellar content of Galactic young massive clusters.}\\
\hline
\hline
Cluster/   &Age   & Mass              & B[e] & O2-6Ia($^+$)/ & O7-9 I/ & LBV/BHG/& YHG & RSG & WN & WC/ & 
Reference \\
Complex    & (Myr)& (10$^3 M_{\odot}$) &      & WN6-9ha      &  B0-9 I & WN9-11h&  &  &  & WO   &  
\\
\hline
\endfirsthead
\caption{continued.}\\
\hline\hline
Cluster/   &Age   & Mass              & B[e] & O2-6Ia($^+$)/ & O7-9 I/ & LBV/BHG/& YHG & RSG & WN & WC/ & 
Reference \\
Complex    & (Myr)& (10$^3 M_{\odot}$) &      & WN6-9ha      &  B0-9 I & WN9-11h&  &  &  & WO   &  
\\
\hline
\endhead
\hline
\endfoot
Pismis 24  & $<1-2$   &    -           &  0   &    1         & 0       & 0     & 0     & 0  & 0  & 1  & 1\\ 
NGC 3603  &  1-2        & $10-16$        &  0   &    5         &   2     & 0     & 0     & 0  & 0  & 0  & 2,3,4 \\
IC 1805     & 1-3   &    -           &  0   &    1         & 1       & 0     & 0     & 0  & 0  & 0  & 5,6\\ 
Westerlund 2        & $<2.5$   & 3              &  0   &    3         &   0     & 0     & 0     & 0  & 0  & 0  & 7,8 \\
NGC 6611 &  $2-3$   &    -        &  0  &    0         & 2       & 0     & 0     & 0  & 0  & 0  & 9,10 \\
Arches     & $2-4$      &   $>10$      & 0    &    28        & 0       & 0     & 0     & 0  & 0  & 0  & 11\\
Mercer 23       & $2-4$    & $3-5$          &  0   &    3         &   5     & 0     & 0     & 0  & 0  & 0  & 12 \\
Havlen-Moffat 1       & $2-4$    &    -           &  0   &    5         & 0       & 0     & 0     & 0  & 0  & 0  & 1\\
DBS2003 179 & $2-5$ & $>7$             &  0   &    8         &   0     & 1       & 0     & 0  & 0  & 0  & 13,14 \\
NGC 6823   & $2-7$    &    -           &  0   &    0         & 5       & 0     & 0      &  0   & 0 & 0 & 5\\
Ruprecht 44& $2.1-3.5$&   -            &  0   &    0         & 0       & 0     & 0     & 0  & 1  & 0  & 1\\
Berkeley 86       &$2.5-3.4$   &    -         &  0   &    0         & 0       & 0     & 0     & 0  & 1  & 0  & 1\\
Cl1806-20  & $3-4$    & $>3$           &  0   &    0         &   4     & 1     & 0     & 0  & 1  & 1  & 15 \\
Glimpse 30 & $3-4.5$  & 3              &  0   &    0         &   0     & 1     & 0     & 0  & 3  & 0  & 16 \\
 VVV CL011  & $3-7$      &    1         & 0    &    1         & 0       & 0    & 0     & 0  & 0  & 0  & 17\\
Quartet    & $3-8$    & $1.3-5.2$      &  0   &    0         &   0     & 2     & 0     & 0  & 0  & 1  & 18 \\
 Berkeley 87       & $\sim3.2$  &    -         &  0   &    0         & 2       & 1?    & 0    & 0  & 0  & 1  & 1\\    
Quintuplet & $3.3-3.6$  &   $>10$      & 0    &    1         & ?       & 8      & 0    & 0  & 2  & 13 & 19,20,21\\
Pismis 20  & $3.6-6.0$ &   -           &  0   &    0         & 1       & 0      & 0    & 0  & 1  & 0  & 1\\
Mercer 81  & $3.7^{+0.4}_{-0.5}$& 10   &  0   &    9         &   0     & 0      & 1    & 0  & 0  & 0  & 22,23 \\
Cl1813-178 & $4-4.5$    & 10           &  0   &    2         &   12     & 1     & 0    & 1  & 3  & 0  & 24 \\
Bochum 7   & $4.3\pm0.5$&    -         &  0   &    0         &     0   & 0      & 0    & 0  & 1  & 0  & 25,26 \\
VVV CL009  & $4-6$      &    1         &  0   &    1         & 5       & 0      & 0    & 0  & 0  & 0  & 17\\
VVV CL074  & $4-6$      &   $>2.5$     & 0    &    1         & 0       & 0      & 0    & 1  & 1  & 1  & 17\\
VVV CL099  & $4-6$      &   $>1.6$     & 0    &    0         & 0       & 0      & 0    & 0  & 1  & 1  & 17\\
NGC 6913   & $4-6$      &     -        &  0   &    0         & 3       & 0     & 0     & 0  & 0  & 0  & 5,27 \\ 
{\bf Westerlund 1} & $\sim5$  &   $\sim100$      & 1    &    0     & $>$100 & 7 & 6    & 4  & 14  & 8 & 25,28,29\\
NGC 6604   & $5\pm2$    &     -        & 0    & 0            & 2 &0 & 0& 0& 0 & 0                     &30,31\\ 
VVV CL036  & $5-7$      &   2.2        & 0    &    0         & 1       & 0      &    0    & 1  & 1  & 0  & 17\\
DBS2003 45 & $5-8$  & $>2$           &  0   &    0         &   6     & 0        & 0    & 0  & 0  & 0  & 32 \\
Gal. Centre&  6         &   $>10$      & 0    &    0         & 27      & 8      & 0    & 0  & 9  & 14 & 33,34\\
NGC 6910    &  $6\pm2$   &     -        & 0    &    0         & 1        &   0   &  0   & 0   & 0 & 0  & 35 \\
Glimpse 20 & $6-8$    & 3.4            &  0   &    0         &   1     & 0      & 1    & 0  & 0  & 1  & 18 \\
NGC 7235   & $6-11$     &   -          &  0   &    0         & 1       & 0     & 0     & 1  & 0  & 0  & 3,36   \\
VVV CL073  & $<7$       &   $>1.4$     & 0    &    1         & 0       & 0      & 0    & 0  & 1  & 0  & 17\\
Markarian 50       & $\sim7.4$  &    -         &  0   &    0         & 1       & 0      & 0    & 0  & 1  & 0  & 1\\
NGC 3293    & $8\pm1$    &    -         &  0   &    0         & 2       & 0      & 0    & 1  & 0   & 0 & 37,38 \\
NGC 2414 &    9          &    -         &  0   &    0         & 1       & 0      & 0    &  0 & 0   & 0 & 39,40 \\
IC 2581    &   10       &    -         &  0   &     0        & 1       & 0      & 0    & 0  & 0   & 0 & 41 \\
Ruprecht 55 & 10        &     -        &  0   &     0        &  1      &  0     &  0   &  0 & 0   & 0  & 39,42\\
NGC 654     & $>10$   &    -         &   0  &    0         & 1       & 0      & 0    & 0  & 0   & 0 & 6, 43\\
NGC 4755   & 11         &    -         &  0   &    0         & 3       & 0      & 0    & 0  & 0  & 0  & 37,44 \\
RSGC 1      & $12\pm2$ & $30\pm10$      &  0   &    0         &   0     & 0      & 1    & 14 & 0  & 0  & 45 \\
Cl1900+14  & $13-15$  & 1              &  0   &    0         &   4     & 0      & 0    & 2  & 0  & 0  & 46\\
$h+{\chi}$ Per & 14   & 20             &  0   &    0         &   15    & 0      & 0    & 6  & 0  & 0  & 47,48 \\
NGC 7419    & $14\pm2$ & 7-10           &  0   &    0         &  0      & 0      & 0    & 5   & 0  & 0  & 49\\ 
Mercer 9        & $15-27$  & $1.6\pm0.4$    &  0   &    0         &   2     & 0      & 0    & 3  & 0  & 0  &50 \\
RSGC 3      & $16-20$  & $30\pm10$      &  0   &    0         &   0     & 0      & 0    & 16 & 0  & 0  & 51,52 \\
RSGC 4      & $16-20$  & $>10$          &  0   &    0         &   0     & 0      & 0    & 9  & 0  & 0  & 52 \\
RSGC 5      & $16-20$  & $>10$          &  0   &    0         &   0     & 0      & 0    & 9  & 0  & 0  &  53\\
RSGC 6      & $16-20$  & $>10$          &  0   &    0         &   0     & 0      & 0    & 8  & 0  & 0  &  54\\
RSGC 2      & $17\pm3$ & $40\pm10$      &  0   &    0         &   0     & 0      & 0    & 26 & 0  & 0  & 55\\   
NGC 457     & 20       &  -             &  0   &    0         &   0     & 0      & 0    & 1  & 0  & 0  & 56,57\\     
NGC 2439    & 20       &   -            &  0   &    0         &   1     & 0      & 0    & 1  & 0  & 0  & 58,59 \\
            &          &                &      &              &         &        &     &    &    &   & \\
{\bf Cyg OB2}     & {\em 2-7}& $>10$          &  1   &    6         &   11    & 2  & 0    & 0  & 2  & 3  & 60,61 \\[1mm]
Danks 1+   & $1.5^{+1.5}_{-0.5}$&$8\pm1.5$& 0 &    5         &   1     & 0   & 0   & 0  & 0  & 0  &  62 \\  
Danks 2+   & $3^{+3}_{-1}$& $3\pm0.8$   &  0   &    0         &   1     & 0   & 0   & 0  & 0  & 1  &  62 \\
G305 field &  -       &  -             &  0   &    1         &   0     & 0    & 0  & 0  & 3  & 4  &  62\\[1mm]
NGC 6231+   & {\em 2.5-5}&  -           &  0   &    0         &   2     & 0    & 0  & 0  & 0  & 1  & 63\\ 
Sco OB1    & {\em 2.5-5}&  -           &  0   &    1         &   8     & 0    & 0  & 0  & 0  & 0  & 64\\[1mm]
Trumpler 14 +      & 1.5      &   -            &  0   &    1         &   0     & 0    & 0  & 0  & 0  & 0  & 1,65\\
Trumpler 16 + & $2-3$    &   -            &  0   &    4         &   3     & 0    & 0  & 0  & 0  & 0  & 1,65\\
Collinder 228 + &         &               &      &                &       &        &  &    &    &    &     \\
Trumpler 15 +      & $6\pm3$  &   -            &  0   &    0         & 1       & 0    & 0  & 0  & 0  & 0  & 1,65\\
Bochum 10      & 7        &   -            &  0   &    0         & 1       & 0    & 0  & 0  & 0  & 1  & 1,65\\[1mm]
W 43        &  -       &  -             &  0   &    3         &   0     & 0    & 0  & 0  & 0  & 0  & 66 \\
NGC 6871    & {\em 2-4}  &    -         &  0   &    0         & 2       & 0    & 0  & 0  & 1  & 0  & 1\\
Trumpler 27       & {\em 3-5} &    -         &  0   &    0         & 7       & 0    & 0  & 1  & 0  & 2  & 1,25\\
   &          &                &      &              &         &        &    &    &   & \\
 {\bf Total}  &          &            &  {\bf 2}     &     {\bf 91}         &  $>${\bf 251}      &  {\bf 31(+1?)}  & {\bf 9}    & {\bf 110}   & {\bf 47}   &  {\bf 54} & \\
\end{longtable}
{Breakdown of the stellar content of known Galactic young stellar aggregates ($\leq$20Myr; 
clusters and OB associations) ordered approximately by increasing age. Clusters/associations
 hosting sgB[e] stars given in bold. The upper pannel consists of clusters, the lower of OB associations (broken 
down where possible  into constituent clusters), non-coeval aggregates (ages given in itallics) and the cluster W43, 
for which no age is given (but its youth is indicated by its associated GH\,{\sc ii} region). NGC3603 appears co-eval with the exception of the two BSGs Sher 23 and 25 which are suggestive of an older ($\sim4$~Myr) poulation.
 Regarding the GC cluster proper for simplicity we  adopt the numbers of stars within the 2 
coherent disc-like structures (Paumard et al. \cite{paumard}) noting that a number of additional  WR candidates 
(but no B[e] stars) have been identified in the surrounds. Classification of the Quintuplet OB supergiants
 is particularly difficult; via comparison to the Arches we reclassify LH099 as WN7-9ha but of the 84 stars Liermann et al. 
(\cite{liermann09}) list as O stars we are unable to determine accurate spectral types and/or luminosity classes.Following the literature convention we restrict YHGs to spectral types A, F and G. Finally, 
we assign a tentative classification of LBV to V439 Cyg; a star with an uncertain history which Polcaro \& Norci (\cite{polcaro}) quote as being identified as 
respectively 
`Mira-like', B3e, a carbon star, B0ep, B1.5~Ve and a candidate LBV over the last $\sim70$yrs.
References used in the construction of this table are
$^1$Massey et al. (\cite{massey}),
$^2$Harayama et al. (\cite{harayama}),
$^3$Melena et al. (\cite{melena}),
$^4$Schnurr et al. (\cite{schnurr}), 
$^5$Massey et al. (\cite{massey95}),
$^6$Shi \& Hu (\cite{shi}),
$^{7,8}$Rauw et al. (\cite{rauw07},\cite{rauw12}), 
$^9$Hillenbrand et al. (\cite{hillenbrand}),
$^{10}$Martayan et al. (\cite{martayan}),
$^{11}$Martins et al. (\cite{martins}),
$^{12}$Hanson et l. (\cite{hanson10}), 
$^{13,14}$Borissova et al. (\cite{borissova08},\cite{borissova12}),
$^{15}$Bibby et al. (\cite{bibby}),
$^{16}$Kurtev et al. (\cite{kurtev}), 
$^{17}$Chen\'{e} et al. (\cite{chene}),
$^{18}$Messineo et al. (\cite{messineo09}), 
$^{19}$Figer et al. (\cite{figer}),
$^{20,21}$Liermann et al. (\cite{liermann09},\cite{liermann12}), 
$^{22}$Davies et al. (\cite{davies12b}), 
$^{23}$de La Fuente et al. (\cite{dlf12}), 
$^{24}$Messineo et al. (\cite{maria11}),
$^{25}$Crowther et al. (\cite{crowther06}), 
$^{26}$Michalska et al. (\cite{michalska}),
$^{27}$Wang \& Hu (\cite{wang00}),
$^{28}$Clark et al. (\cite{cncg05}),
$^{29}$ Negueruela et al. (\cite{negueruela10}), 
$^{30}$Barbon et al. (\cite{barbon}),
$^{31}$De Becker et al. (\cite{deb}), 
$^{32}$Zhu et al. (\cite{zhu}), 
$^{33}$Paumard et al. (\cite{paumard}), 
$^{34}$Martins et al. (\cite{martins07}),
$^{35}$Kolaczkowski et al. (\cite{kol}),
$^{36}$Sowell (\cite{sowell}),
$^{37}$Evans et al. (\cite{evans05}),
$^{38}$Baume et al. (\cite{baume}),
$^{39}$Kharchenko et al. (\cite{kharch}),
$^{40}$Fitzgerald et al. (\cite{f79}),
$^{41}$Turner (\cite{turner}),
$^{42}$Bosch et al. (\cite{bosch}),
$^{43}$Pandey et al. (\cite{panday}),
$^{44}$Aidelman et al. (\cite{aidelman}),
$^{45,46}$Davies et al. (\cite{davies08},\cite{davies09}), 
$^{47}$Slesnick et al. (\cite{slesnick}),
$^{48}$Currie et al. (\cite{currie}), 
$^{49}$Marco \& Negueruela (\cite{marco}),
$^{50}$Messineo et al. (\cite{messineo10}), 
$^{51}$Clark et al. (\cite{clark09}), 
$^{52}$Negueruela et al. (\cite{neg10b}), 
$^{53}$Negueruela et al. (\cite{negueruela11}), 
$^{54}$Gonz\'{a}lez-Fern\'{a}ndez \& Negueruela (\cite{gonzalez}),
$^{55}$Davies et al. (\cite{davies07a}), 
$^{56}$Zhang et al. (\cite{zhang}),
$^{57}$Fitzsimmons et al. (\cite{fitzs}),
$^{58}$Ramsay \& Pollacco (\cite{ramsay}),
$^{59}$Kaltcheva et al. (\cite{kaltcheva}),
$^{60}$Hanson (\cite{hanson03}), 
$^{61}$Negueruela et al. (\cite{negueruela08}),
$^{62}$Davies et al. (\cite{davies12}), 
$^{63}$Sana et al. (\cite{sana}), 
$^{64}$Ankay et al. (\cite{ankay}), 
$^{65}$Smith (\cite{smith06}),
$^{66}$Blum et al. (\cite{blum}).}
}

\begin{thebibliography}{}
\bibitem[2012]{aidelman}
Aidelman, Y., Cidale, L. S., Zorec, J. \& Airas, M. L. 2012, A\&A, 544, A64
\bibitem[2001]{ankay}
Ankay, A., Kaper, L., de Bruijne, J. H. J. et al. 2001, A\&A, 370, 170 
\bibitem[1998]{app}
Appenzeller, I., Fricke, J, F\"urtig, W. et al.\ 1998, The Messenger, 94, 1
\bibitem[2010]{ardila}
Ardila, D. R., Van Dyk, S. D., Makowiecki, W., et al. 2010, ApJS, 191, 301
\bibitem[2012]{aret}
Aret, A., Kraus, M., Muratore, M. F. \& Borges Fernandes, M. 2012, \mnras, 423, 284
\bibitem[2000]{barbon}
Barbon, R., Carraro, G., Munari, U., Zwitter, T. \& Tomasella, L.
2000, A\&AS, 144, 451
\bibitem[2013]{bartlett}
Bartlett, E. S., Clark J. S., Coe, M. J., Garcia, M. R. \&
Uttley, P. 2013, MNRAS, 
429, 1213
\bibitem[2003]{baume}
Baume, G., V\'azquez, R. A., Carraro, G. \& Feinstein, A. 2003, A\&A, 402, 549    
\bibitem[2008]{bibby}
Bibby, J. L., Crowther, P. A., Furness, J. P. \& Clark, J. S. 2008, MNRAS, 386, L23
\bibitem[2006]{bik}
 Bik, A., Kaper, L. \& Waters, L. B. F. M.   2006, A\&A, 455, 561
\bibitem[1999]{blum}
Blum, R. D., Damineli, A. \& Conti, P. S. 1999, AJ, 117, 1392
\bibitem[2010]{boersma}
Boersma, C., Bauschlicher, C. W., Allamandola, L. J. et al. 2010, A\&A, 511, A32
\bibitem[2007]{bonanos07}
Bonanos, A.Z.\ 2007, \aj, 133, 2696 
\bibitem[2009]{bonanos09}
Bonanos, A. Z., Massa, D. L., Sewilo, M. et al. 2009, AJ, 138, 1003 
\bibitem[1970]{bks70}
Borgman, J., Koornneef, J., Slingerland, J.\ 1970, \aap, 4, 248
\bibitem[2008]{borissova08}
Borissova, J., Ivanov, V. D., Hanson, M. M., et al. 2008, A\&A, 488, 151
\bibitem[2012]{borissova12}
Borissova, J., Georgiev, L., Hanson, M. M., et al. 2012, A\&A, 546, 110 
\bibitem[2003]{bosch}
Bosch, G., Barba, R., Morrell, N. et al.2003, MNRAS, 341, 169
\bibitem[1997]{brandner}
Brandner, W., Grebel, E. K., Chu, Y.-H. \& Weis, K. 1997, ApJ, 475, L45
\bibitem[2013]{chene}
Chen\'{e} A.-N., Borissova, J., Bonatto, C.,  et al. 2013, A\&A, 549, 98 
\bibitem[2006]{chiar}
Chiar, J. E. \& Tielens, A. G. G. M.\ 2006, \apj, 637, 774
\bibitem[1998]{clark98}
Clark, J. S., Fender, R. P., Waters, L. B. F. M., et al.\ 1998, \mnras, 299, 43
\bibitem[2000]{clark00}
Clark, J. S., Miroshnichenko, A. S., Larionov, V. M., et al.\ 2000, \aap, 356, 50
\bibitem[2005]{cncg05}
Clark, J. S., Negueruela, I., Crowther, P. A. \& Goodwin, S.\ 2005, \aap, 434, 949  
\bibitem[2008]{clark08}
Clark, J. S., Muno, M. P., Negueruela, I., et al.\ 2008, \aap, 477, 147
\bibitem[2009]{clark09}
Clark, J. S., Negueruela, I., Davies, B., et al.\ 2009, \aap, 498, 109
\bibitem[2010]{clark10}
Clark, J. S., Ritchie, B. W. \& Negueruela, I.\ 2010, \aap, 514, A87
\bibitem[2011]{clark11a}
Clark, J. S., Ritchie, B. W., Negueruela, I., et al.\ 2011, \aap, 531, A28
\bibitem[2013]{clark13}
Clark, J. S., Bartlett, E. S., Coe, M. J. et al. 2013, \aap, in press (arXiv:1305.0459C)
\bibitem[2012]{cottaar}
Cottaar, M., Meyer, M. R., Andersen, M. \& Espinoza, P.\ 2012, \aap, 539, A5
\bibitem[2006a]{bsg}
Crowther, P. A., Lennon, D. J. \& Walborn, N. R. 2006a, A\&A, 446, 279
\bibitem[2006b]{crowther06}
Crowther, P. A., Hadfield, L. J., Clark, J. S., Negueruela, I. \& Vacca, W. D.\ 2006b, \mnras, 372, 1407
\bibitem[2010]{currie}
Currie, T., Hernandez, J., Irwin, J. et al. 2010, ApJS, 186, 191
\bibitem[2001]{danchi}
Danchi, W. C., Tuthill, P. G. \& Monnier, J. D.\ 2001, \apj, 562, 440
\bibitem[1997]{davidson}
Davidson, K. \& Humphreys, R. M.\ 1997, \araa, 35, 1
\bibitem[2007a]{davies07a}
Davies, B., Figer, D. F., Kudritzki, R.-P., et al. 2007a, ApJ, 671, 781
\bibitem[2007b]{davies}
Davies, B., Oudmaijer, R. D. \& Sahu, K. C.\ 2007b, \apj, 671, 2059
\bibitem[2008]{davies08}
Davies, B., Figer, D. F., Law, C. J., et al. 2008, ApJ, 676, 1016
\bibitem[2009]{davies09}
Davies B. Figer, D. F., Kudritzki, R.-P., et al. 2009, ApJ, 707, 844
\bibitem[2012a]{davies12} 
Davies, B., Clark, J. S., Trombley, C. et al\ 2012a, \mnras, 419, 1871
\bibitem[2012b]{davies12b} 
Davies, B., de La Fuente, D., Najarro, F., et al. 2012b, MNRAS, 419, 1860 
\bibitem[2005]{deb}
De Becker, M., Rauw, G. Blomme, R. et al. 2005, A\&A, 437, 1029
\bibitem[2000]{detal00}
Dekker, H., D'Odorico, S., Kaufer, A., Delabre, B. \& Kotzlowski, H.\ 2000, SPIE, 4008, 534
\bibitem[2012]{dlf12}
de La Fuente, D., Najarro, F., Davies, B., Figer, D. F., 2012, arxiv:1210.1781
\bibitem[2013]{doran}
Doran, E. I., Crowther, P. A., de Koter, A. et al. 2013, A\&A, 558, A134
\bibitem[2010]{dougherty}
Dougherty, S. M., Clark, J. S., Negueruela, I., Johnson, T. \& Chapman, J. M.\ 2010, \aap, 511, A58
\bibitem[2001]{dullemond}
Dullemond, C. P., Dominik, C. \& Natta, A.\ 2001, \apj, 560, 957
\bibitem[2012]{dunstall}
Dunstall, P. R., Fraser, M., Clark, J. S. et al. 2012, A\&A, 542, 50 
\bibitem[2002]{eggenberger}
Eggenberger, P., Meynet, G. \& Maeder, A. 2002, A\&A, 386, 576
\bibitem[2005]{evans05}
Evans, C. J., Smartt, S. J., Lee, J.-K., et al. 2005, A\&A, 437, 467 
\bibitem[2011]{evans}
Evans, C. J., Taylor, W. D., H\'{e}nault-Brunet, V., et al.\ 2011, \aap, 530, A108
\bibitem[1999]{figer}
Figer, D. F., McLean, I. S. \&  Morris, M.\ 1999, \apj, 514, 202
\bibitem[2004]{filliatre}
Filliatre, P. \& Chaty, S.\ 2004, \apj, 616, 469
\bibitem[1979]{f79}
Fitzgerald, M. P., Luiken, M., Maitzen, H. M. \& Moffat, A. F. J. 1979, A\&AS, 37, 345 
\bibitem[1993]{fitzs}
Fitzsimmons, A. 1993, A\&AS, 99, 15
\bibitem[2006]{fuchs}
Fuchs, Y., Koch Miramond, L. \& \'{A}brah\'{a}m, P. 2006, A\&A, 445, 1041
\bibitem[1996]{degraauw}
de Graauw, T., Haser, L. N., Beintema, D. A., et al.\ 1996, 315, L49
\bibitem[2012]{gonzalez}
Gonz\'{a}lez-Fern\'{a}ndez, C. \& Negueruela, I. 2012, A\&A, 539, A100
\bibitem[1985]{gosset}
Gosset, E., Hutsemekers, D., Surdej, J. \& Swings, J. P. 1985, A\&A, 153, 71
\bibitem[1980]{grandi80}
Grandi, S. A.\ 1980, \apj, 238, 10
\bibitem[2012]{graus}
Graus, A. S., Lamb, J. B. \& Oey, M. S. 2012, \apj, 759, 10
\bibitem[2006]{groh06}
Groh, J. H., Damineli, A., Teodoro, M. \& Barbosa, C.L.\ 2006, \aap, 457, 591
\bibitem[2010]{groh10}
Groh, J. H., Nielsen, K. E., Damineli, A. et al.\ 2011, \aap, 517, A9
\bibitem[2013]{groh13}
Groh, J. H., Meynet, G., Georgy, C. \& Ekstrom, S. 2013, A\&A, in press 
(2013arXiv1308.4681)
\bibitem[2007]{grundstrom}
Grundstrom, E. D., Gies, D. R., Hillwig,  T. C., et al.\ 2007, \apj, 667, 505 
\bibitem[1995]{gummersbach}
Gummersbach, C. A., Zickgraf, F.-J. \& Wolf, B. 1995, A\&A, 302, 409
\bibitem[2012]{gvara}
Gvaramadze, V. V. \& Menten, K. M. 2012, A\&A, 541, A7
\bibitem[1986]{hs86}
Hamann, F. \& Simon, M.\ 1986, \apj, 311, 909
\bibitem[1988]{hs88}
Hamann, F. \& Simon, M.\ 1988, \apj, 327, 867
\bibitem[2003]{hanson03}
Hanson, M. M. 2003, ApJ, 597, 957
\bibitem[2010]{hanson10}
Hanson, M. M., Kurtev, R., Borissova, J., et al. 2010, A\&A 516, A35
\bibitem[2008]{harayama}
Harayama, Y. Eisenhauer, F. \& Martins, F. 2008, ApJ, 675, 1319
\bibitem[1993]{hillenbrand}
Hillenbrand, L. A., Massey, P. Strom, S .E. \& Merrill, K. M. 1993, AJ, 106, 1906
\bibitem[2006]{hillier}
Hillier, D.J.\ 2006, in: Stars with the B[e] phenomenon, ASP Conf. Ser. 355, 39
\bibitem[2002]{hofmann}
Hofmann, K.-H., Balega, Y., Ikhsanov, N. R., Miroshnichenko, A. S. \& Weigelt, G.\ 2002, \aap, 395, 891
\bibitem[1996]{vdh}
van de Hucht, K. A., Morris, P. W., Williams, P. M., et al.\ 1996, \aap, 315, L193
\bibitem[2005]{hump05}
Humphreys, R. M., Davidson, K., Ruch, G. \& Wallerstein, G.\ 2005, \aj, 129, 492
\bibitem[2002]{hynes}
Hynes, R. I., Clark, J. S., Barsukova, E. A., et al.\ 2002, \aap, 392, 991
\bibitem[2001]{kaltcheva}
Kaltcheva, N., Gredel, R. \& Fabricius, C. 2001, A\&A, 372, 95
\bibitem[2006]{kastner}
Kastner, J. H., Buchanan, C. L., Sargent, B. \& Forrest, W. J.\ 2006, \apj, 638, L29
\bibitem[2010]{kastner10}
Kastner, J. H., Buchanan, C., Saghai, R., Forrest, W. J. \& Sargent, B. A.\ 2010, \aj, 139, 1993
\bibitem[2005]{kharch}
Kharchenko, N. V., Piskunov, A. E., Roeser, S., Schilbach, E. \& Scholz, R. D. 2005, A\&A, 438, 1163
\bibitem[2004]{kol}
Kolaczkowski, Z., Pigulski, A., Kopacki, G. \& Michalska, G.
2004, AcA, 54, 33
\bibitem[2002]{kraemer}
Kraemer, K. E., Sloan, G. C., Price, S. D., Walker, H. J. 2002, \apjs, 
140, 349 
\bibitem[2009]{kraus09}
Kraus, M. 2009, A\&A, 494, 253
\bibitem[2007]{kraus07}
Kraus, M., Borges Fernandes, M. \& de Ara\'{u}jo, F. X. 2007, \aap, 463, 627
\bibitem[2010]{kraus10}
Kraus, M., Borges Fernandes, M. \& de Ara\'{u}jo, F. X. 2010, \aap, 517, A30
\bibitem[2013]{kraus12}
Kraus, M., Oksala, M., Nickeler, D., et al. 2013, \aap, 549, A28
\bibitem[2012]{skraus}
Kraus, S., Calvet, N., Hartmann, L. et al. 2012, ApJ, 746, L2
\bibitem[2007]{kurtev}
Kurtev, R., Borissova, J., Georgiev, L., Ortolani, S. \& Ivanov, V. D. 2007, A\&A, 475, 209
\bibitem[1984]{kwan84}
Kwan, J.\ 1984, \apj, 283, 70
\bibitem[1996]{lamers96}
Lamers, H. J. G. L. M., Morris, P.W., Voors, R. H. M., et al.\ 1996, \aap, 315, 225
\bibitem[1998]{lamers}
Lamers, H. J. G. L. M., Zickgraf, F.-J., de Winter, D., Houziaux, L. \& Zorec, J.\ 1998, \aap, 340, 117 
\bibitem[1998]{langer98}
Langer, N. \& Heger, A.  1998, Astrophys. Space Sci. Libr. 233, 235
\bibitem[2003]{langer03}
Langer, N., Wellstein, S. \& Petrovic, J.\ 2003, in: IAU Symp. 212, A Massive Star Odyssey, from Main Sequence to Supernova, eds. van der Hucht, K.A., Herrero, A., \& Esteban, C., 275
\bibitem[2009]{liermann09}
 Liermann, A., Hamann, W.-R. \& Oskinova, L. M. 2009, A\&A, 494, 1137
\bibitem[2010]{liermann}
Liermann, A., Kraus, M., Schnurr, O. \& Fernandes, M. 2010, \mnras, 408, L6
\bibitem[2012]{liermann12}
Liermann, A., Hamann, W.-R. \& Oskinova, L. M., 2012, A\&A, 540, A14
\bibitem[2003]{lobel03}
Lobel, A., Dupree, A. K., Stefanik, R. P. et al.\ 2003, \apj, 583, 923
\bibitem[1998]{lutz98}
Lutz, D., Kunze, D., Spoon, H. W.W . \& Thornley, M. D.\ 1998, \aap, 333, L75
\bibitem[2013]{marco}
Marco, A. \& Negueruela, I. 2013, A\&A, 552, A92
\bibitem[2007]{marcolino}
Marcolino, W. L. F., de Arua\'{u}jo, F. X., Lorenz-Martins, S. \& Borges Fernandes, M. 2007, \aj, 133, 489 
\bibitem[2008]{martayan}
Martayan, C., Floquet, M., Hubert, A. M. et al. 2008, A\&A, 489, 459 
\bibitem[2007]{martins07}
Martins, F., Genzel, R., Hillier, D. J., et al. 2007, A\&A, 468, 233
\bibitem[2008]{martins}
Martins, F., Hillier, D. J., Paumard, T. et al.\ 2008, \aap, 478, 219
\bibitem[1995]{massey95}
Massey, P., Johnson, K. E.,  \&  DeGioia-Eastwood, K.\ 1995, ApJ, 454, 141
\bibitem[2000]{massey00}
Massey, P., Waterhouse, E. \&  DeGioia-Eastwood, K.\ 2000, \aj, 119, 2214
\bibitem[2001]{massey}
Massey, P., DeGioia-Eastwood, K. \& Waterhouse, E.\ 2001, \aj, 121, 1050
\bibitem[2008]{melena}
Melena, N. W., Massey, P., Morrell, N. I. \& Zangari, A. M. 2008,
AJ, 135, 878
\bibitem[2007]{mengel}
Mengel, S. \& Tacconi-Garman, L. E.\ 2007, \aap, 466, 151
\bibitem[2009]{messineo09}
Messineo, M., Davies, B., Ivanov, V. D., et al. 2009, ApJ, 697, 701
\bibitem[2010]{messineo10}
Messineo, M., Figer, D. F., Davies, B., et al. 2010, ApJ, 708, 1241
\bibitem[2011]{maria11}
Messineo, M., Davies, B., Figer, D. F., et al.\ 2011, \apj, 733, 41
\bibitem[2013]{michalska}
Michalska, G., Niemczura, E., Pigulski, A. Ste\'{s}licki, M. \&
Williams, A. 2013, MNRAS, 429, 1354
\bibitem[2007]{miro07}
Miroshnichenko, A. S.\ 2007, \apj, 667, 497
\bibitem[2007]{moon}
Moon, D.-S., Kaplan, D. L., Reach, W. T., et al.\ 2007, \apj, 671, L53
\bibitem[1996]{morris}
Morris, P. W., Eenens, P. R. J., Hanson, M. M., Conti, P. S. \& Blum, R. D. 1996, ApJ, 470, 597
\bibitem[2006]{muno}
Muno, M. P., Clark, J. S., Crowther, P. A., et al.\ 2006, \apj, 636, L41
\bibitem[2005]{negueruela05}
Negueruela, I. \& Clark, J. S.\ 2005, \aap, 436, 541
\bibitem[2008]{negueruela08}
Negueruela, I., Marco, A., Herrero, A. \& Clark, J. S. 2008, A\&A, 487, 575
\bibitem[2010a]{negueruela10}
Negueruela, I., Clark, J. S. \& Ritchie, B. W.\ 2010a, \aap, 516, A78
\bibitem[2010b]{neg10b}
Negueruela, I., Gonz\'{a}lez-Fern\'{a}ndez, C., Marco, A., Clark, J. S., Mart\'{i}nez-N\'{u}\~{n}ez, S.\ 2010b, \aap, 513, A74
\bibitem[2011]{negueruela11}
Negueruela, I., Gonz\'{a}lez-Fern\'{a}ndez, C., Marco, A. \& Clark, J. S. 2011, A\&A, 528, A59
\bibitem[1997]{okazaki}
Okazaki, A. 1997, A\&A, 318, 548
\bibitem[2012]{oksala}
Oksala, M. E., Kraus, M., Arias, M. L. et al. 2012, MNRAS, 426, L56
\bibitem[1998]{oud98b}
Oudmaijer, R. D., Proga, D., Drew, J. E. \& de~Winter, D.\ 1998, \mnras, 300, 170
\bibitem[2005]{panday}
Pandey, A. K., Upadhyay, K., Ogura, K. et al. 2005, MNRAS, 358, 1290
\bibitem[2002]{pasq02}
Pasquini, L., Avila, G., Blecha, A., et al.\ 2002, The Messenger, 110, 1
\bibitem[2006]{paumard}
Paumard T., Genzel, R., Martins, F. et al. 2006, ApJ, 643, 1011 
\bibitem[2005]{petrovic}
Petrovic, J., Langer, N. \& van der Hucht, 
K.A.\ 2005, \aap, 435, 1013 
\bibitem[2006]{pod}
Podsiadlowski, P., Morris, T. S. \& Ivanova, N.\ 2006, in: Stars with the B[e] phenomenon, ASP Conf. Ser. 355, 259
\bibitem[1998]{polcaro}
Polcaro, V. F. \& Norci, L.\ 1998, \aap, 339, 75
\bibitem[1992]{ramsay}
Ramsay, G. \& Pollacco, D. L. 1992, A\&AS, 94, 73
\bibitem[2007]{rauw07}
Rauw, G., Manfroid, J. Gosset, E., et al. 2007, A\&A, 463, 981
\bibitem[2012]{rauw12}
 Rauw, G., Sana, H., Naz\'{e}, Y.  2011, A\&A, 535, A40 
\bibitem[2011]{richardson}
Richardson, N. D., Gies, D. R. \& Williams, S. J. 2011, \aj, 142 201
\bibitem[2009a]{ritchie09a}
Ritchie, B. W., Clark, J. S., Negueruela, I. \& Crowther, P.A.\ 2009a, \aap, 507, 1585
\bibitem[2009b]{ritchie09b}
Ritchie, B. W., Clark, J. S., Negueruela, I. \& Najarro, F.\ 2009b, \aap, 507, 1597
\bibitem[2010a]{ritchie10}
Ritchie, B. W., Clark, J. S., Negueruela, I. \& Langer, N.\ 2010a, \aap, 520, A48
\bibitem[2010b]{ritchie10b}
Ritchie, B. W., Clark,   J. S. \& Negueruela, I.\ 2010b, Soci\'{e}t\'{e} Royale des Sciences de
  Li\`{e}ge, Bulletin, vol. 80, p. 628-633 (Proceedings of the 39th
  Li\'{e}ge Astrophysical Colloquium, 12-16 July 2010, edited by G.
  Rauw, M. De Becker, Y. Naz\'{e}, J.-M. Vreux \& P. Williams)
\bibitem[2013]{ritchie13}
Ritchie, B. W., Clark, J. S., Negueruela, I., et al.\ 2013, \aap, in prep.
\bibitem[2006]{sana}
Sana, H., Gosset, E., Rauw, G., Sung, H. \&  Vreux, J.-M. 2006, MNRAS, 372, 661 
\bibitem[2008]{schnurr}
Schnurr, O., Casoli, J., Chen\'{e}, A.-N., Moffat, A. F. J. \& 
St-Louis, N. 2008, MNRAS, 389, L38
\bibitem[2000]{sheikina}
Sheikina, T. A., Miroshnichenko, A. S., \& Corporan, P. 2000, in The Be Phenomenon in Early-Type Stars, IAU Colloquium 175, Astronomical
Society of the Pacific Conference Series, Vol. 214, edited by Myron A. Smith and Huib F. Henrichs, page 494. 
\bibitem[1999]{shi}
Shi, H. M. \&  Hu, J. Y. 1999, A\&AS, 136, 313
\bibitem[2003]{sigut}
Sigut, T. A. A \& Pradhan, A. K.\ 2003, \apjs, 145, 15 
\bibitem[2006]{skinner}
Skinner, S. L., Simmons, A. E., Zhekov, S. A., et al.\ 2006, \apj, 639, L35
\bibitem[2002]{slesnick}
Slesnick, C. L., Hillenbrand, L. A., Massey, P., 2002, ApJ, 576, 880
\bibitem[2003]{sloan}
Sloan, G. C., Kraemer, K. E., Price, S. D. \& Shipman, R. F.\ 2003, \apjs, 147, 379
\bibitem[2002]{smartt}
Smartt, S. J., Lennon, D. J., Kudritzki, R. P. et al. 2002, A\&A, 391, 979
\bibitem[2006]{smith06}
Smith, N. 2006, MNRAS, 367, 763
\bibitem[2002]{smith02}
Smith, N., Gehrz, R. D., Stahl, O., Balick, B. \& Kaufer, A.\ 2002, \apj, 578, 464
\bibitem[2007]{smith07}
Smith, N., Bally, J., Walawender, J. 2007, AJ, 134, 846
\bibitem[2011]{smith11}
Smith, N., Gehrz, R.D., Campbell, R. et al\ 2011, MNRAS, 418, 1959
\bibitem[1987]{sowell}
Sowell, J. R. 1987, ApJS, 64 241
\bibitem[1996]{st96}
Stothers, R. B. \& Chin, C.-W.\ 1996, \apj, 468, 842
\bibitem[1973]{turner}
Turner, D. G. 1973, AJ,78, 597
\bibitem[2000]{wang00}
Wang, J.-J. \& Hu, J.-Y. 2000, A\&A, 356, 118
\bibitem[2012]{wang}
Wang, Y., Weigelt, G., Kreplin, A. et al. 2012, \aap, 545, L10
\bibitem[1999]{wellstein}
Wellstein, S. \& Langer, N.\ 1999, \aap, 350, 148
\bibitem[2008]{weintroub}
Weintroub, J., Moran, J. M., Wilner, D. J., et al.\ 2008, \apj, 677, 1140
\bibitem[1961]{w61}
Westerlund, B. E.\ 1961, \pasp, 73, 51
\bibitem[1987]{w87}
Westerlund, B. E.\ 1987, \aaps, 70, 311
\bibitem[2012]{wheelwright}
Wheelwright, H. E., de Wit, W. J., Weigelt, G., Oudmaijer, R. D. \& Ilee, J. D.
2012, \aap, 543, A77
\bibitem[1985]{white}
White, R. L. \& Becker, R. H. 1985, \apj, 297, 677
\bibitem[2012]{zhang}
Zhang, X. B., Luo, C.Q. \& Fu, J. N. 2012, AJ, 144, 86
\bibitem[2009]{zhu}
Zhu, Q., Davies, B., Figer, D. F. \& Trombley 2009, ApJ, 702, 929
\bibitem[2003]{zick03}
Zickgraf, F.-J.\ 2003, \aap, 408, 257
\bibitem[1985]{zick85}
Zickgraf, F.-J., Wolf, B., Stahl, O., Leitherer, C. \& Klare, G.\ 1985, \aap, 143, 421
\bibitem[1986]{zick86}
Zickgraf, F.J., Wolf, B., Stahl, O., Leitherer, C. \& Appenzeller, I. 1986, \aap, 163, 119
\bibitem[1989]{zick89}
Zickgraf, F.-J., Wolf, B., Stahl, O. \& Humphreys, R. M. 1989, \aap, 220, 206

\end{thebibliography}
\end{document}